\documentclass[journal=enfuem,manuscript=article,layout=twocolumn]{achemso}

\usepackage[version=3]{mhchem} 
\usepackage{amsmath,amssymb}
\usepackage[dvipsnames]{xcolor}

\makeatletter
\let\l@addto@macro\relax
\makeatother
\usepackage[fontsize=9pt]{scrextend}

\usepackage{graphicx}
\usepackage{dcolumn}
\usepackage{bm}
\usepackage[utf8]{inputenc}
\usepackage[T1]{fontenc}
\usepackage{mathptmx}
\usepackage{etoolbox}
\usepackage{bigints}
\usepackage[dvipsnames]{xcolor}
\usepackage{comment} 
\usepackage{tabularray}
\usepackage{multirow}

\author{A. Borrero}
\affiliation{Laboratorio de Simulaci\'on Molecular y Qu\'imica Computacional, CIQSO-Centro de Investigaci\'on en Qu\'imica Sostenible and Departamento de Ciencias Integradas, Universidad de Huelva, 21006 Huelva Spain}

\author{A. Díaz-Acosta}
\affiliation{Laboratorio de Simulaci\'on Molecular y Qu\'imica Computacional, CIQSO-Centro de Investigaci\'on en Qu\'imica Sostenible and Departamento de Ciencias Integradas, Universidad de Huelva, 21006 Huelva Spain}

\author{S. Blazquez}
\affiliation{Departamento de Química Física, Facultad de Ciencias Químicas, Universidad Complutense de Madrid, 28040 Madrid, Spain}

\author{I. M. Zerón}
\affiliation{Laboratorio de Simulaci\'on Molecular y Qu\'imica Computacional, CIQSO-Centro de Investigaci\'on en Qu\'imica Sostenible and Departamento de Ciencias Integradas, Universidad de Huelva, 21006 Huelva Spain}

\author{J. Algaba}
\affiliation{Laboratorio de Simulaci\'on Molecular y Qu\'imica Computacional, CIQSO-Centro de Investigaci\'on en Qu\'imica Sostenible and Departamento de Ciencias Integradas, Universidad de Huelva, 21006 Huelva Spain}

\author{M. M. Conde}
\email{maria.mconde@upm.es}
\affiliation{Departamento de Ingeniería Química Industrial y Medio Ambiente, Escuela Técnica Superior de Ingenieros Industriales, Universidad Politécnica de Madrid, 28006, Madrid, Spain}

\author{F. J. Blas}
\email{felipe@uhu.es}
\affiliation{Laboratorio de Simulaci\'on Molecular y Qu\'imica Computacional, CIQSO-Centro de Investigaci\'on en Qu\'imica Sostenible and Departamento de Ciencias Integradas, Universidad de Huelva, 21006 Huelva Spain}

\let\oldmaketitle\maketitle
\let\maketitle\relax

\title[Phase behavior of CO$_{2}$ hydrates in salty water]{Three-phase equilibria of CO$_2$ hydrate from computer simulation in presence of NaCl}

\begin{document}
\begin{tocentry}
\centering
\includegraphics[angle=0,width=6.4cm]{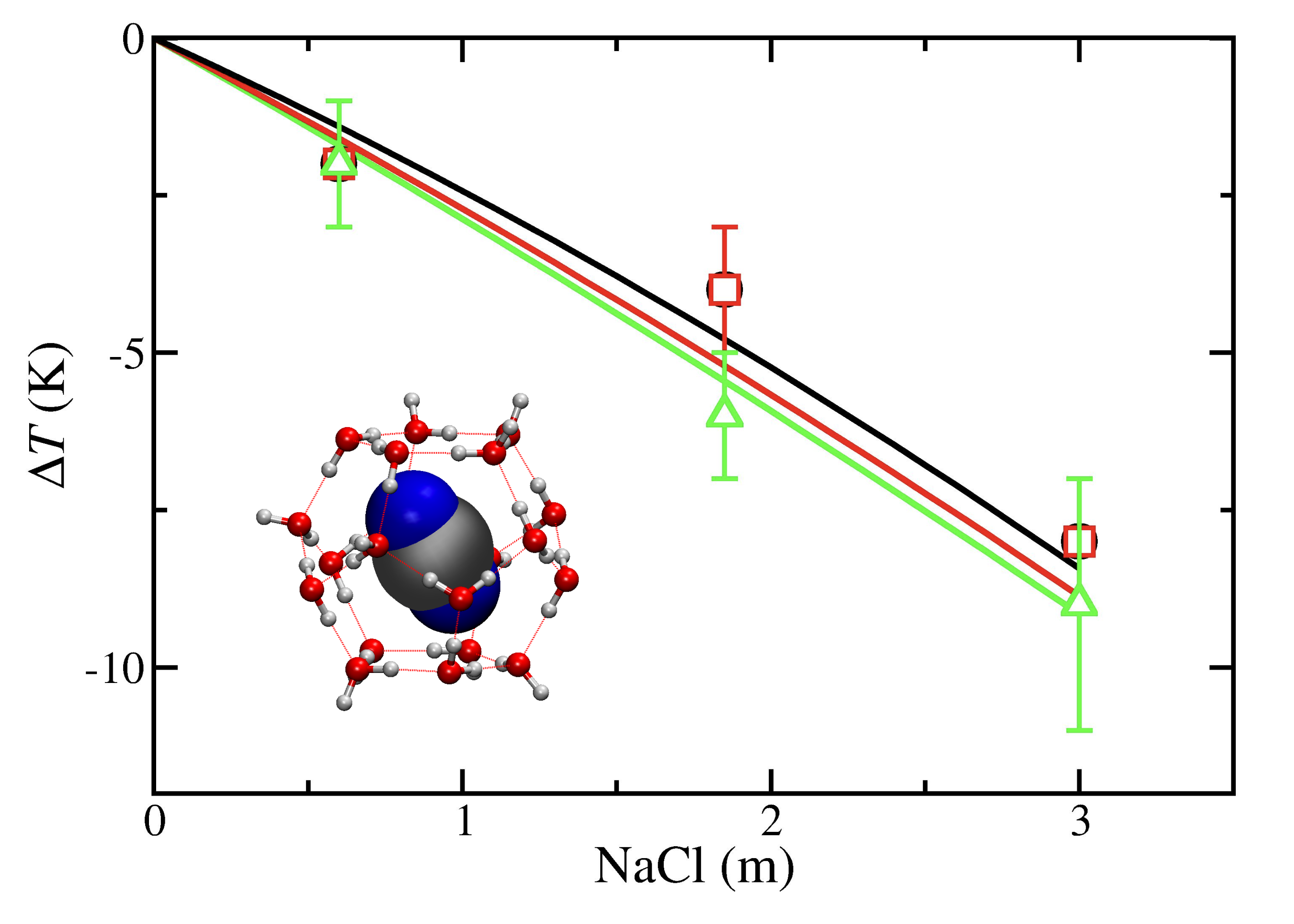}

Cryoscopic decrease effect of a CO$_{2}$ hydrate at different pressures.
\end{tocentry}
\maketitle
%
\twocolumn[
\begin{@twocolumnfalse}
\oldmaketitle

\begin{abstract}
In this work, the cryoscopic decrease effect, as a function of the NaCl concentration, on the carbon dioxide (CO$_2$) hydrate dissociation line conditions has been determined through molecular dynamic simulations. In particular, we have determined the three-phase (solid hydrate--aqueous phase--liquid CO$_2$) coexistence temperature at 100, 400, and 1000\,bar at several initial NaCl concentrations in the aqueous phase, from 0.0 to $3.0\,\text{m}$, using the direct-coexistence technique. We have used the well-known TIP4P/2005 and TraPPe force fields for water and CO$_2$ molecules, respectively. Also, the water-salt interactions are described using the Madrid-2019 force field, which has been specifically developed for various salts in combination with the TIP4P/2005 water model. According to the results obtained in this work, the dissociation temperature of the CO$_2$ hydrate decreases when the NaCl concentration in the initial aqueous phase increases. The results obtained are in excellent agreement with experimental data reported in the literature. We have also observed how the dynamic of melting and growth of the CO$_2$ hydrate becomes slower when the NaCl concentration is increased. As a consequence, larger simulation times (in the order of dozens of microseconds) are necessary when the NaCl concentration increases. Finally, we have also analyzed finite-size effects on the three-phase coexistence temperature of these systems by performing simulations, at 400\,bar,  with two different system sizes at two different NaCl concentrations (0.0 and $3.0\,\text{m}$). Non-negligible deviations have been found between the results obtained from both system sizes.
\end{abstract}
\end{@twocolumnfalse}
]


%

%

\section{Introduction}
Clathrates are non-stoichiometric crystalline inclusion compounds in which small guest molecules, such as methane (CH$_{4}$), carbon dioxide (CO$_{2}$), hydrogen (H$_{2}$), and nitrogen (N$_{2}$), and medium guest molecules, such as ethane (C$_2$H$_6$), propane (C$_3$H$_8$), iso-butane (C$_4$H$_{10}$), and tetrahydrofuran (THF), among numerous other species, are enclathrated within the voids left by a periodic arrange of hydrogen-bonded particles (host).~\cite{Sloan2007a,Ripmeester2022a} When the network of hydrogen-bonded molecules is built with
water molecules (H$_2$O), clathrates are  called
clathrate hydrates or simply hydrates.~\cite{Sloan2007a,Ripmeester2022a}  In the last decades, hydrates have been the subject of fundamental and applied research~\cite{Sloan2007a,Ripmeester2022a,Ripmeester2016a,Ratcliffe2022a} because of their promising applications in CO$_2$ capture,~\cite{Ma2016a,Dashti2015a,C5CP07202F,Cannone2021a,Duc2007a,Choi2022a,Lee2014a,Yang2014a,Ricaurte2014a,Kvamme2007a} H$_2$ storage and transport,~\cite{veluswamy2014hydrogen,lee2005tuning,Brumby2019a} N$_2$ recovery from industrial emissions,~\cite{Yi2019a,Hassanpouryouzband2018a} H$_2$S as hydrate promoter,~\cite{liang2010crystal} and  also  from an energetic point of view since there is more CH$_4$ trapped in hydrates in the nature than in conventional fossil fuel reservoirs.~\cite{Lee2001a,Ruppel2017a} Recent estimates predict the existence of 3000 trillion cubic meters of natural gas in the form of hydrates. This is much larger than the 600-850 trillion cubic meters available from conventional sources.~\cite{Chong2016}  A large amount of CH$_4$ hydrate reservoirs have been known for decades in different parts of the world.~\cite{Kvenvolden1988a,Koh2012a} However, only in recent years, countries such as the USA and Japan have initiated pilot plants to extract CH$_4$ from seabed hydrates.~\cite{Sloan2007a} Obviously, the efficient and optimal exploitation of CH$_4$ from these resources requires, among many other disciplines and technology development, a precise knowledge of its thermodynamics, including phase equilibrium, as well as interfacial and dynamic properties.

As it has been mentioned previously, from an environmental point of view, hydrates are also valuable and strategic materials in the field of CO$_2$ capture\cite{Ma2016a,Dashti2015a,Cannone2021a,Duc2007a,Choi2022a,Lee2014a,Yang2014a,Ricaurte2014a} and storage.~\cite{Kvamme2007a} The increase of the amount of CO$_2$ in the Earth's atmosphere, mainly due to human industrial activity, is one of the main sources of climate change and radically affects our society on a planetary scale. The possibility of using sI-hydrates for CO$_2$ storage has been considered for decades due to their thermodynamic stability under relatively mild conditions.~\cite{Sloan2007a,Ripmeester2022a} In this context, it is of great interest the possibility of recovering CH$_4$ from ocean floor CH$_4$ hydrates at the same time that CO$_2$ is stored in situ. \cite{10.1063/5.0155143,10.1063/5.0179655} This is possible since CO$_2$ hydrates are more stable than CH$_4$ hydrates at natural gas reservoirs' pressure and temperature conditions on the seafloor. However, in order to collect CH$_4$ from seafloor hydrates at the same time that CO$_2$ is stored in situ, it is necessary to obtain a precise knowledge of the thermodynamic, phase equilibrium, interfacial, and dynamic properties of CH$_4$ and CO$_2$ hydrates, not only at the temperature and pressure conditions of the seafloor but also at the salinity conditions of seawater. For this reason, it is essential to locate the so-called dissociation line or three-phase coexistence line, hydrate - aqueous phase - CO$_2$/CH$_4$, in a pressure-temperature phase diagram at the salinity conditions of seawater. Precise knowledge of the three-phase coexistence line provides information about the pressure and temperature conditions at which the hydrate is stable. The use of molecular dynamic simulations to study the effect of the NaCl concentration on the CH$_4$ hydrate dissociation conditions has already been reported in the literature.  Yagasaki \emph{et al.}\cite{Yagasaki2014a} performed molecular dynamic simulations at $m=0.6$ and $4.8\,\text{kg/mol}$ to study the dissociation kinetics of the CH$_4$ hydrate at three different temperatures over the dissociation one. They find that the NaCl could have both deceleration and acceleration effects on the
kinetics of hydrate dissociation. In particular, they observed an acceleration effect at the $m=4.8\,\text{kg/mol}$ NaCl concentration, which corresponds to the saturated conditions at ambient temperature and pressure, and temperatures close to the CH$_4$ hydrate dissociation one. Very recently, Blazquez \emph{et al.}~\cite{Blazquez2023b} have determined the three-phase coexistence temperature for a methane hydrate system in equilibrium with a NaCl solution and a methane gas phase. The 
study has been performed at $400\,\text{bar}$ and a range of concentration given in molality, below $4\,\text{m}$. In this work, we concentrate on the determination of the three-phase CO$_{2}$ hydrate-aqueous solution-CO$_{2}$-rich liquid phase or dissociation line with different modalities up to $3\,\text{m}$.

Seawater is a complex aqueous solution made up of many components, including a large number of ions. However, most of the physical properties of seawater are essentially dependent on a single parameter: salinity, i.e., the concentration of ions in water. Indeed, salinity is one of the fundamental properties of seawater and is key for understanding biological and physical processes in the oceans. Absolute salinity is defined as the ratio of the mass of dissolved material in seawater to the mass of seawater\cite{Unesco1980} (in grams per kilogram of seawater). The salinity of the oceans is typically between 31 and 38\,g/kg. According to the International Association for the Physical Sciences of the Oceans (IAPSO),~\cite{Culkin1998a,Millero2008} the reference composition of standard seawater is defined in terms of the molar fractions of its 15 major constituents, including water, with the main constituent being sodium chloride (NaCl).~\cite{Holland1978a,Millero2008} In particular, sodium and chloride ions account for up to 90\% of the total electrolytes dissolved in water. Therefore, in most studies, seawater is modeled as an aqueous solution of NaCl with a salinity of about 35\,g/kg. This concentration is often expressed in terms of molality, i.e., the number of moles of NaCl per kilogram of solvent (water). In the case of seawater, a NaCl salinity of 35\,g/kg is equivalent to a NaCl molality of 0.6\,m.

From an experimental point of view, there are several studies in the literature that have determined the dissociation line of the hydrate of CO$_2$ not only in the presence of seawater but also at higher salinity conditions, reaching molalities up to 5\,m.~\cite{Dholabhai1993a,Englezos1994a,Tohidi1997a,Bozzo1975a,Duan2006} In particular, experimental studies show that the presence of NaCl in water causes the well-known cryoscopic lowering or melting point depression of the CO$_2$ hydrate. In particular, the dissociation temperature of this hydrate, when it is in contact with a NaCl 0.6\,m aqueous solution, is approximately 2\,K lower than when the NaCl molality of the aqueous solution is zero.~\cite{Dholabhai1993a,Duan2006} The dissociation temperature decreases when the amount of NaCl in the aqueous phase increases. Unfortunately, molecular simulation studies to determine the effect of seawater salinity on the dissociation line of CO$_2$ hydrate are scarce. One of the few found is the recent work of Vlugt and coworkers about the microscopic behavior of CO$_2$ hydrates in oceanic sediments\cite{mi2024molecular}. At this point, an interesting question arises: are the molecular models available in the literature able to reproduce the cryoscopic decrease of the CO$_2$ hydrate dissociation line due to the presence of dissolved salts in water through molecular simulations? In a very recent work, Blazquez \emph{et al.}~\cite{Blazquez2024a} have determined the solubility of CO$_2$ in salty water through molecular dynamic simulations. In particular, they found that the combination of the TIP4P/2005,\cite{Abascal2005a} Madrid-2019,\cite{Zeron2019,10.1063/5.0077716} and TraPPE-UA\cite{Potoff2001a} models for water, NaCl-water interactions, and CO$_2$, reproduces accurately the density of the aqueous phase as well as the solubility and the salting out effect of the CO$_2$ in salty water. This combination of molecular models used in the work of Blazquez \emph{et al.}~\cite{Blazquez2024a} is far from being arbitrary. The TIP4P/2005\cite{Abascal2005a} water and TraPPE-UA\cite{Potoff2001a} CO$_2$ models have been used previously by Miguez \emph{et al.}~\cite{Miguez2015a} to determine the dissociation temperature of the CO$_2$ hydrate. Although in the work of Miguez \emph{et al.}~\cite{Miguez2015a} they conclude that the results predicted by the mentioned model combination were about $\approx30-45$\,K lower than the experimental ones, they were able to reproduce qualitatively the dissociation line of the CO$_2$ hydrate. In the same work, they conclude that the TIP4P/Ice\cite{Abascal2005b} water model provides better predictions for the dissociation condition of the CO$_2$ hydrate. Unfortunately, the Madrid-2019\cite{Zeron2019,10.1063/5.0077716} NaCl model has been specifically designed for the TIP4P/2005\cite{Abascal2005a} water model. For this reason, in this work, we have decided to use the TIP4P/2005 instead of the TIP4P/Ice water model since we are more interested in analyzing accurately the cryoscopic lowering effect on the CO$_2$ hydrate than in determining quantitatively the CO$_2$ hydrate dissociation temperature in salty water.

The organization of this paper is as follows: In Sec. II, we describe the simulation details and the methodology. The results obtained in this work at the different thermodynamic conditions and NaCl concentration are discussed in detail in Sec. III. Finally, conclusions are presented in Sec. IV.

\section{Methodology}

\subsection{Simulation details and molecular models}

In this work, all molecular dynamics simulations have been carried out using GROMACS (version 4.6.5, double precision).~\cite{VanDerSpoel2005a} Our system is composed of three components: water, CO$_2$, and NaCl. It is essential to describe their interactions accurately. Water molecules have been modeled using the well-known rigid and non-polarizable TIP4P/2005\cite{Abascal2005a} model. For the CO$_2$ molecules, we employed the TraPPE~\cite{Potoff2001a} (Transferable Potentials for Phase Equilibria) force field. This model has been successfully tested for the phase equilibria of CO$_2$ hydrates \cite{Blazquez2024b,Algaba2024a,Algaba2024b}. To describe the water-salt interactions in our system, we have used the Madrid-2019 force field.\cite{Zeron2019,10.1063/5.0077716} This force field has been developed for various salts in combination with TIP4P/2005, assigning partial charges to the ions (e.g., $\pm$0.85$e$ for Na$^+$ and Cl$^-$).

It is worth noting that the combination of the TIP4P/2005 water model and the TraPPE CO$_2$ model has previously been used to study the three-phase coexistence line of CO$_2$ hydrate in the absence of salt.~\cite{Miguez2015a} These results showed significant deviations from experimental data available in the literature. In their work, Miguez {\it et al.}~\cite{Miguez2015a} extended their study by also using the TIP4P/Ice~\cite{Abascal2005b} model, which resulted in a remarkable improvement in their results. This improvement is not coincidental: Conde and Vega~\cite{Conde2013a} previously demonstrated that, for methane hydrates, there is a correlation between the melting temperature of ice I$_h$ ($T_m$) and the three-phase coexistence temperature ($T_3$) of the methane hydrate. Models that accurately predict the melting point of ice I$_h$, such as TIP4P/Ice, are better suited to reproduce the experimental hydrate equilibrium line.

However, the goal of the present work is to determine the three-phase equilibrium of CO$_2$ hydrate in the presence of salt. Recent studies have demonstrated that using a model with partial charges improves the prediction of the equilibrium line at both intermediate and high salt concentrations, as shown for systems involving ice I$_h$,~\cite{Lamas2022a,YAN2023123198} methane hydrate,~\cite{Blazquez2023b} and even the role of static electric fields in seawater icing.~\cite{ZHAO2025126744} These findings support our decision to use the Madrid-2019 force field with partial charges, which, in combination with the TIP4P/2005 model, has demonstrated significant superiority in describing a wide range of properties of aqueous solutions compared to unit charge models \cite{Blazquez2020a,Blazquez2024a,Blazquez2023a,sedano2022maximum,blazquez2023computation,10.1063/5.0077716}. In fact, when using other force fields for the study of the cryoscopic decrease of the \textit{T$_3$} of methane hydrate such as Joung-Cheatham \cite{JC} or Smith and Dang \cite{SmithDang} models, the effect of the salt model has been negligible at low concentrations. \cite{Fernandez-Fernandez2019a,Blazquez2023b}

When salt is added to the aqueous phase, the thermodynamic conditions for the dissociation of the CO$_2$ hydrate are modified. The three-phase coexistence temperature shifts depending on the amount of NaCl added to the aqueous solution phase. Specifically, we are interested in the deviation of $T_3$ as a function of the molality of NaCl in the aqueous solution, rather than the absolute value of $T_3$. Our aim is to study the cryoscopic effect on freezing-point depression as a function of NaCl concentration. For this reason, we employed the Madrid-2019 force field, where the non-bonded interaction parameters of NaCl have been optimized for the TIP4P/2005 water model. A summary of the non-bonded interaction parameters of the molecular models is presented in Table \ref{Model}. The Lennard-Jones (LJ) parameters for interactions not listed in Table \ref{Model} are determined using the Lorentz-Berthelot rules. Note that the Lorentz-Berthelot rule is not applied to water-ion or ion-ion interactions in the Madrid-2019 force field.

\begin{table}
\caption{Non-bonded interaction parameters of water, CO$_2$ and NaCl models used in this work. Water molecules are described using the TIP4P/2005 model.~\cite{Abascal2005a} For NaCl, we use the parameters from Madrid-2019 force field,~\cite{Zeron2019,10.1063/5.0077716} and LJ interaction parameters for CO$_2$ are taken from Ref. \cite{Potoff2001a}
}
\centering
\begin{tabular}{lccccccc}
\hline\hline
Atom & & $q$(e)  & &$\sigma(\text{\AA})$ & & $\varepsilon(\text{kJ/mol})$&\\
\hline
\multicolumn{8}{c}{Water} \\
\hline
O$_{\text{w}}$           & & --  & &3.1589 & &0.7749 &\\
H & & +0.5564 & & --  & & -- &\\
M  & & -1.1128 &  & -- & & -- &\\
\hline
\multicolumn{8}{c}{CO$_2$} \\
\hline
C & & +0.70 & & 2.80 & & 0.224478 &\\
O & & -0.35 & & 3.05 & & 0.656806 & \\
\hline
\multicolumn{8}{c}{NaCl} \\
\hline
Na$^+$ & & +0.85 & & 2.21737 & & 1.472356 &\\
Cl$^-$ & & -0.85 & & 4.69906 & & 0.076923 &\\
Na$^+$-- Cl$^-$ & & -- & & 3.00512 & & 1.438894 &\\
Na$^+$-- O$_{\text{w}}$  & & -- & & 2.60838 & & 0.793388 &\\
Cl$^-$-- O$_{\text{w}}$  & & -- & & 4.23867 & & 0.061983 &\\
\hline\hline
\end{tabular}
\label{Model}
\end{table}

All molecular dynamic simulations have been carried out in the $NPT$ ensemble. In order to avoid any stress from the solid CO$_2$ hydrate structure, the volume fluctuations are performed independently in each space direction. The three-phase coexistence temperature, $T_3$, has been estimated at three different pressures (100, 400, and $1000\,\text{bar}$) and several NaCl concentrations (0, 0.6, 1.85, and $3.0\,\text{m}$). The $T_3$ value at each pressure and concentration has been obtained by using the direct-coexistence technique.~\cite{Conde2010a,Miguez2015a,Costandy2015a,Michalis2015a} Following this approach, the hydrate phase, the aqueous solution phase (with different NaCl concentrations), and the liquid CO$_2$ phase are put together in the same simulation box. By keeping constant the pressure, $P$, and performing simulations at different temperatures, $T$, it is possible to evaluate the temperature at which the three phases coexist in equilibrium, $T_3$. If $T>T_3$ then, the initial three-phase system evolves to a two-phase system since the hydrate phase melts releasing water and CO$_2$ molecules to the aqueous solution and liquid CO$_2$ phases. Contrary, if $T<T_3$ then the hydrate phase will grow. Here it is important to take into account that, due to the presence of NaCl in the aqueous solution, the hydrate phase will not grow until extinguishing the aqueous solution phase since the concentration of NaCl in the aqueous solution increases at the same time that the molecules of water move from the aqueous solution to the hydrate phase. As a consequence, the system evolves from the initial configuration to a hydrate phase in equilibrium with the liquid CO$_2$ phase and, also, in equilibrium with a supersaturated NaCl aqueous solution phase. We consider that $T<T_3$ if a couple of extra slabs of hydrates are formed even if we do not observe the complete crystallization of the aqueous solution phase. The $T_3$ value at a given initial NaCl concentration is obtained as the middle temperature between the highest temperature at which the hydrate phase grows and the lowest temperature at which the hydrate phase melts. 


In order to maintain constant temperature and pressure, the v-rescale thermostat\cite{Bussi2007a} and the anisotropic Parrinello-Rahman barostat\cite{Parrinello1981a} are used, both with a time constant of 2 ps. For the Parrinello-Rahman barostat, a compressibility value of $4.5\times10^{-5}\,\text{bar}^{-1}$ is applied anisotropically in all three directions of the simulation box. Dispersive and Coulombic interactions are truncated using a cutoff value of $1.0\,\text{nm}$ in both cases. No long-range corrections or modifications to the Lorentz-Berthelot combining rule are applied for the dispersive interactions between water and CO$_{2}$ molecules. For the Coulombic interactions, long-range particle mesh Ewald (PME) corrections~\cite{Essmann1995a} are employed with a mesh width of $0.1\,\text{nm}$ and a relative tolerance of $10^{-5}$.

\begin{table*}
\caption{ Initial number of water, CO$_2$, and NaCl molecules in the three phases involved in the equilibrium for the L and S configurations at each NaCl concentration.}
\label{molecules}
\begin{tabular}{c c c c c c c c c c c c c c c c c}

\hline
\hline
\multirow{2}{*}{Configuration} &  \multirow{2}{*}{NaCl concentration (m)} & \multicolumn{3}{c}{Hydrate phase} & &
\multicolumn{3}{c}{Aqueous phase} & &
\multicolumn{1}{c}{Liquid CO$_2$ phase} & \\
     \cline{3-5} \cline{7-9} \cline{11-12}
      & & Unit Cell  & Water & CO$_2$ & & Water  & Na$^+$ & Cl$^-$ & & CO$_2$  &  \\
      \hline
      \multirow{4}{*}{L} & 0 &\multirow{4}{*}{3x3x3}&\multirow{4}{*}{1242}&\multirow{4}{*}{216}& &\multirow{4}{*}{2484}&0&0&&\multirow{4}{*}{1000} &\\
       & 0.6 &&&& &&27&27&& &\\
       & 1.85 &&&& &&83&83&& &\\
       & 3.0 &&&& &&134&134&& &\\
      \hline
      \multirow{2}{*}{S} & 0 &\multirow{2}{*}{2x2x2}&\multirow{2}{*}{368}&\multirow{2}{*}{64}& &\multirow{2}{*}{1110}&0&0&&\multirow{2}{*}{320} &\\
       & 3.0 &&&& &&60&60&& &\\
      \hline
      \hline
\end{tabular}
\end{table*}

\subsection{Box size}

In this work, we used two types of simulation box sizes: a smaller box labeled as S and a larger box labeled as L. Specifically, we chose the S-size configuration to compare with the results for $T_3$ from the original work without salt performed by Miguez \emph{et al.}~\cite{Miguez2015a}. They studied the three-phase coexistence line of CO$_2$ hydrate at different pressures using the same potential parameters for water and CO$_2$ as in the present work. The initial simulation box used in the work of Miguez \emph{et al.}~\cite{Miguez2015a} was built by replicating the CO$_2$ hydrate unit cell twice in each space direction ($2\times2\times2$). The total number of molecules of water and CO$_2$ in the hydrate phase was 368 and 64, respectively. Then, the same number of molecules of water was used to the aqueous phase, and, finally, 192 molecules of CO$_2$ were placed in the liquid CO$_2$ phase. We followed the same protocol to build the hydrate phase of the initial S configuration in this work. However, we increased the number of molecules in the aqueous phase and the liquid CO$_2$ phase to 1110 and 320 water and CO$_2$ molecules, respectively. This adjustment was necessary because we now include NaCl molecules in the aqueous phase. If the aqueous phase is too small and $T<T_3$, when the hydrate phase grows and the molecules of water move from the aqueous solution to the hydrate phase, the concentration of NaCl increases rapidly avoiding the expansion of the hydrate phase. Thus, we have increased the number of water molecules in the initial aqueous solution by a $\thicksim3$ factor in the S configuration compared to the original work of Miguez \emph{et al.}~\cite{Miguez2015a}.

As demonstrated recently by some of us,~\cite{Blazquez2024b,Algaba2024a,Algaba2024b}, the $T_3$ value determined using the direct coexistence technique is not free of finite-size effects. Therefore, it is crucial to use large enough systems to accurately determine the impact of the NaCl concentration on the $T_3$ value. In order to avoid finite-size effects, we have built the large initial simulation box (L) by replicating the CO$_2$ hydrate unit cell three times in each space direction ($3\times3\times3$). The number of water and CO$_2$ molecules in the initial hydrate phase is 1242 and 216, respectively. We have also increased the number of water and CO$_2$ molecules in both the aqueous solution and the liquid CO$_2$ phases. In the case of the aqueous solution phase, we doubled the number of water molecules relative to the hydrate phase (2484 water molecules in the aqueous solution). Additionally, we have placed 1000 CO$_2$ molecules in the liquid CO$_2$ phase. As it has been demonstrated recently,~\cite{Blazquez2024b,Algaba2024a} this arrangement regarding the number of molecules in the simulation box prevents the formation of a liquid CO$_2$ drop within the aqueous solution. The formation of a liquid drop or a gas bubble leads to an incorrect determination of the $T_3$, allowing the growth of the hydrate phase at temperatures above the real $T_3$ value. A summary of the number of molecules in each phase for the two configurations (S and L) and the initial NaCl concentrations in the aqueous phase for each configuration can be found in Table \ref{molecules}.

\section{Results}

In this section, we present the different $T_3$ results obtained as a function of the NaCl concentration in the aqueous phase and the pressure. We also analyze the effect of the simulation box size on the $T_3$ values at $400\,\text{bar}$ when the initial NaCl concentration is 0 and $3.0\,\text{m}$. Also, we analyze the effect of the NaCl concentration on the dynamic of growth and melting of the CO$_2$ hydrate. Finally, we compare the cryoscopic decrease effect, as a function of the NaCl concentration, on the CO$_2$ hydrate obtained from simulation with experimental data taken from the literature.

\subsection{NaCl $0.0\,\text{m}$}

\begin{figure}
\centering
\includegraphics[width=0.9\columnwidth]{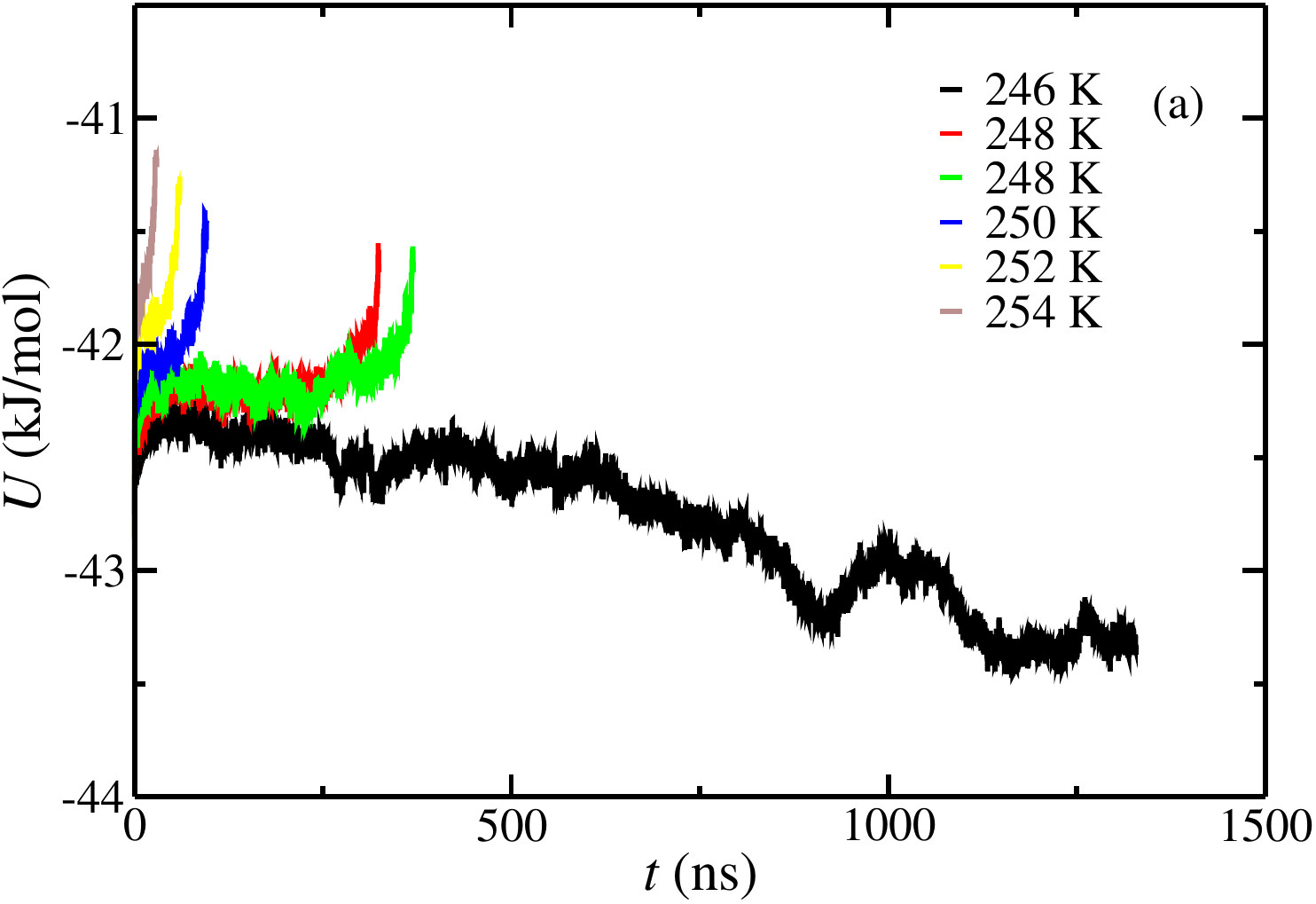}
\includegraphics[width=0.9\columnwidth]{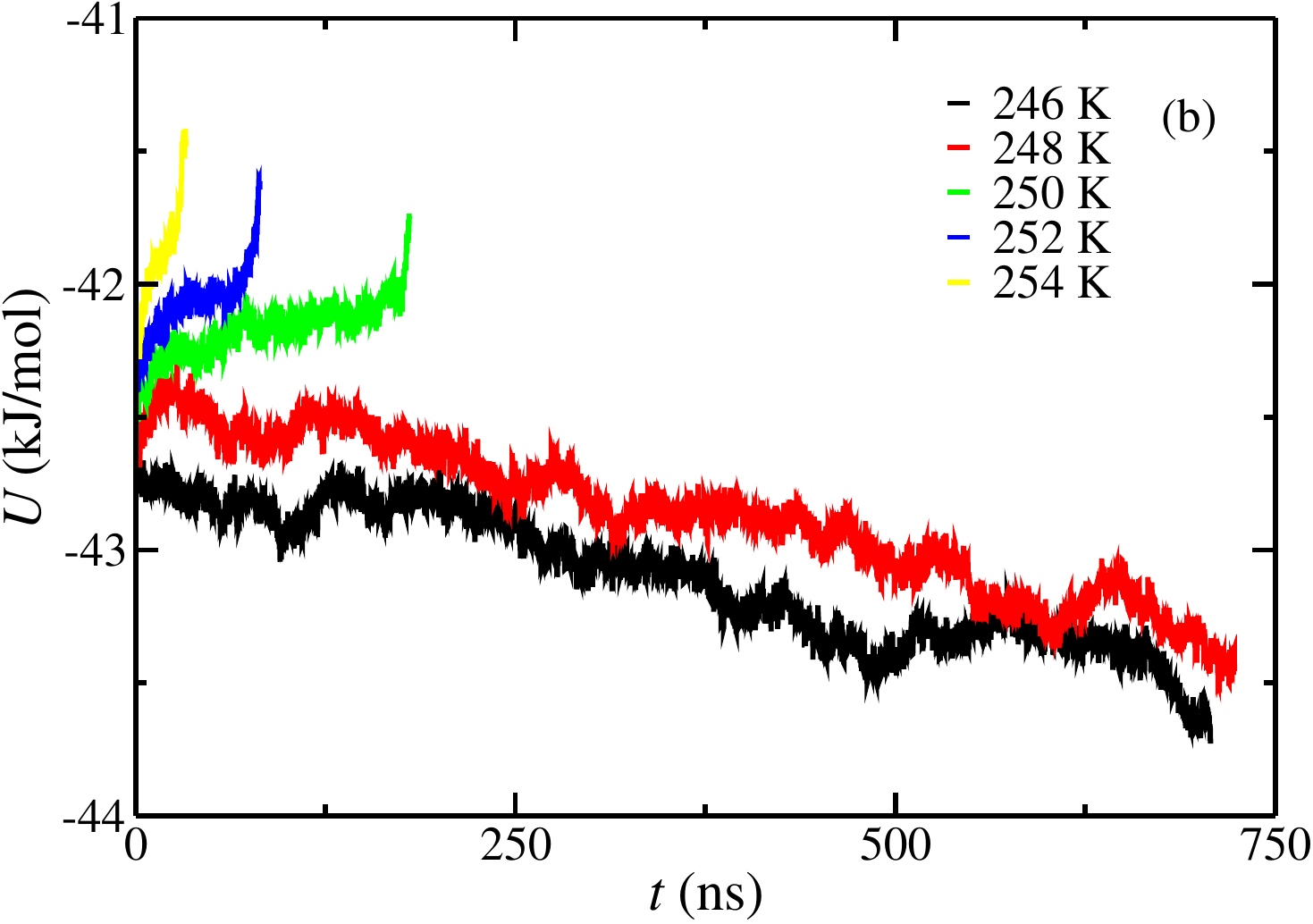}
\includegraphics[width=0.9\columnwidth]{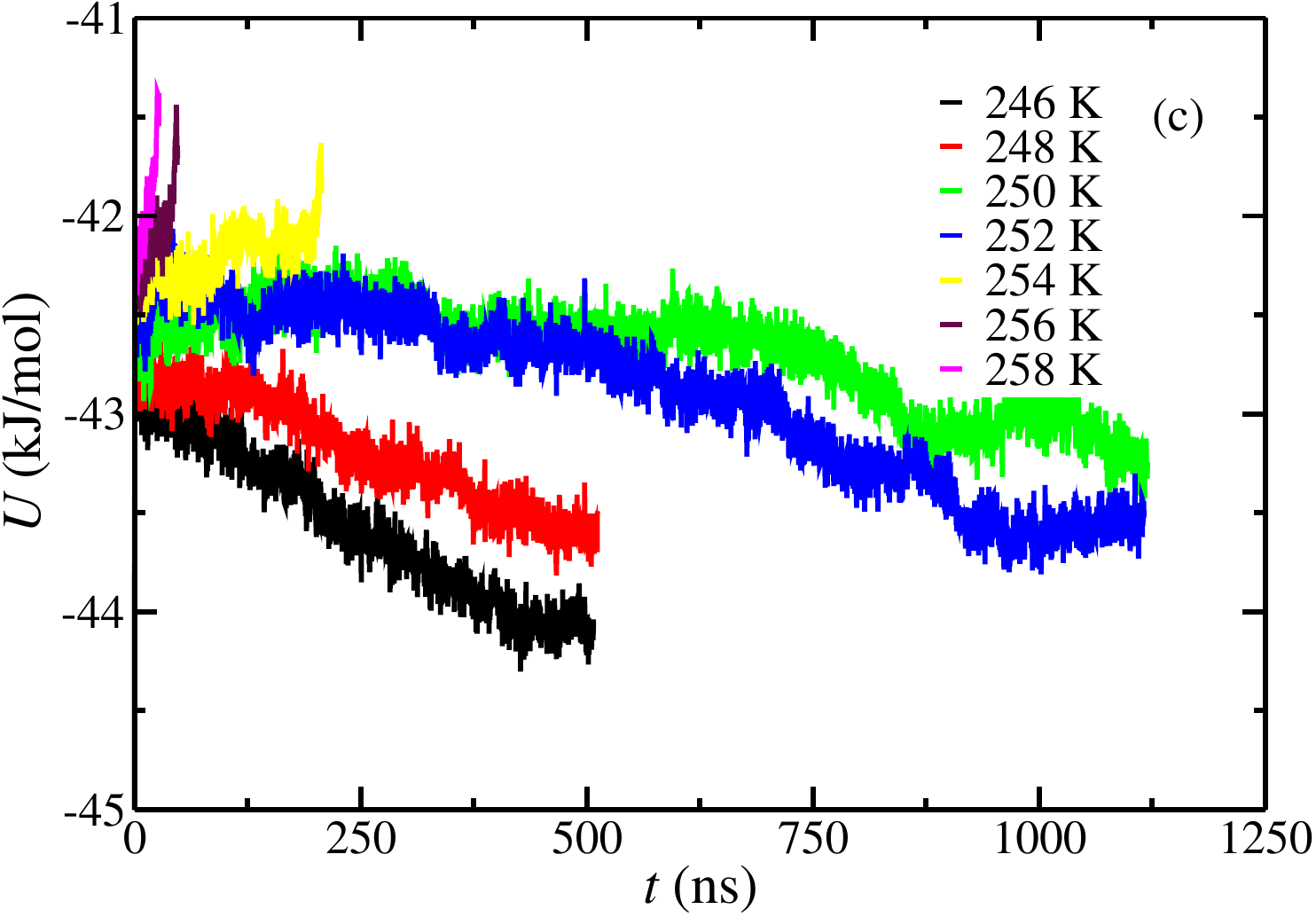}
\caption{Potential energy as a function of time as obtained from molecular dynamic simulations at different temperatures and pressures using configuration L simulation box. Figures (a), (b), and (c) correspond to results obtained at 100, 400, and $1000\,\text{bar}$ respectively. In all cases, the initial aqueous phase is a pure water phase without NaCl. The temperatures studied at each pressure are represented in the legend.}
\label{CO2_simulations_sinsal}
\end{figure}

First of all, we determine the three-phase coexistence temperature, $T_3$, at 100, 400, and $1000\,\text{bar}$ without the presence of NaCl in the aqueous solution phase using configuration L, and at a $400\,\text{bar}$ using configuration S. Since we are going to determine the change in the $T_3$ value as a function of the NaCl concentration, we need to know the exact $T_3$ value without the presence of NaCl. As it has been mentioned previously, in the work of Miguez \emph{et al.},~\cite{Miguez2015a} they determined the $T_3$ of the CO$_2$ hydrate using, as well as in this work, the TIP4P/2005\cite{Abascal2005a} and TraPPE\cite{Potoff2001a} models to describe the molecules of water and CO$_2$ respectively. However, the total number of molecules used by Miguez \emph{et al.}~\cite{Miguez2015a}  is different than that used in this work. As has been demonstrated by some of us in previous work,~\cite{Algaba2024a} finite-size effects on the $T_3$ determination of the CO$_2$ hydrate becomes negligible when non-stoichiometric configurations with larger unit cells, like $3\times3\times3$ and $4\times4\times4$, are used. It means that the $T_3$ values obtained by using configuration L and S can be different from that obtained in the original work of Miguez \emph{et al.}~\cite{Miguez2015a} Since we are interested in determining how the $T_3$ of the CO$_2$ hydrate is affected by the NaCl concentration in the aqueous phase, first, it is necessary to determine accurately the $T_3$ value with both simulation boxes used in this work. For this reason, the $T_3$ values at 100, 400, and $1000\,\text{bar}$ have been calculated using configuration L and, also, the $T_3$ at $400\,\text{bar}$ has been calculated using configuration S. In both cases, the initial aqueous solution phase was formed only by molecules of water.

The potential energy as a function of the simulation time obtained from configuration L  at 100, 400, and $1000\,\text{bar}$ are represented in Fig. \ref{CO2_simulations_sinsal}. The $T_3$ values obtained using configuration L at 100 (a), 400 (b), and 1000 (c) bar are 247(1), 249(1), and $253(1)\,\text{K}$ respectively. As can be observed, the $T_3$ values obtained in this work by using configuration L and the results obtained by Miguez \emph{et al.}~\cite{Miguez2015a} are the same within the error bars (249(2), 249(2), and $252(2)\,\text{K}$ at 100, 400, and $1000\,\text{bar}$ respectively).  As we have pointed out previously, the size and the number of molecules in configuration L is much higher than that used in the work of Miguez \emph{et al.}~\cite{Miguez2015a}, so one could expect different results. However, it is important to remark that the cutoffs employed in both works are different. In the work of Míguez \emph{et al.}~\cite{Miguez2015a}, simulations were carried out using a cutoff value of $0.9\,\text{nm}$ and homogeneous long-range corrections for the Lennard-Jones (LJ) dispersive interactions, while in this work the value of the cutoff is $1.0\,\text{nm}$ and we have not employed long-range corrections for the LJ interactions. Some of us have demonstrated in a very recent work~\cite{Algaba2024b} the effect of the cutoff and long-range corrections on the hydrate $T_3$ determination. The use of homogeneous long-range corrections increases the $T_3$ of the system. However, one should be careful when they are applied to a system where different phases coexist since, technically, they can be only applied to homogeneous systems. In this particular case, the use of homogeneous long-range corrections compensates for the small size of the system used in the original work of Miguez \emph{et al.}~\cite{Miguez2015a} This explains why the results obtained in this work using configuration L are similar to those obtained by Míguez \emph{et al.}~\cite{Miguez2015a} even when the size of the simulation boxes are different. A summary of the results obtained in this work is presented in Table \ref{T3}.

\begin{figure}
\centering
\includegraphics[width=1.0\columnwidth]{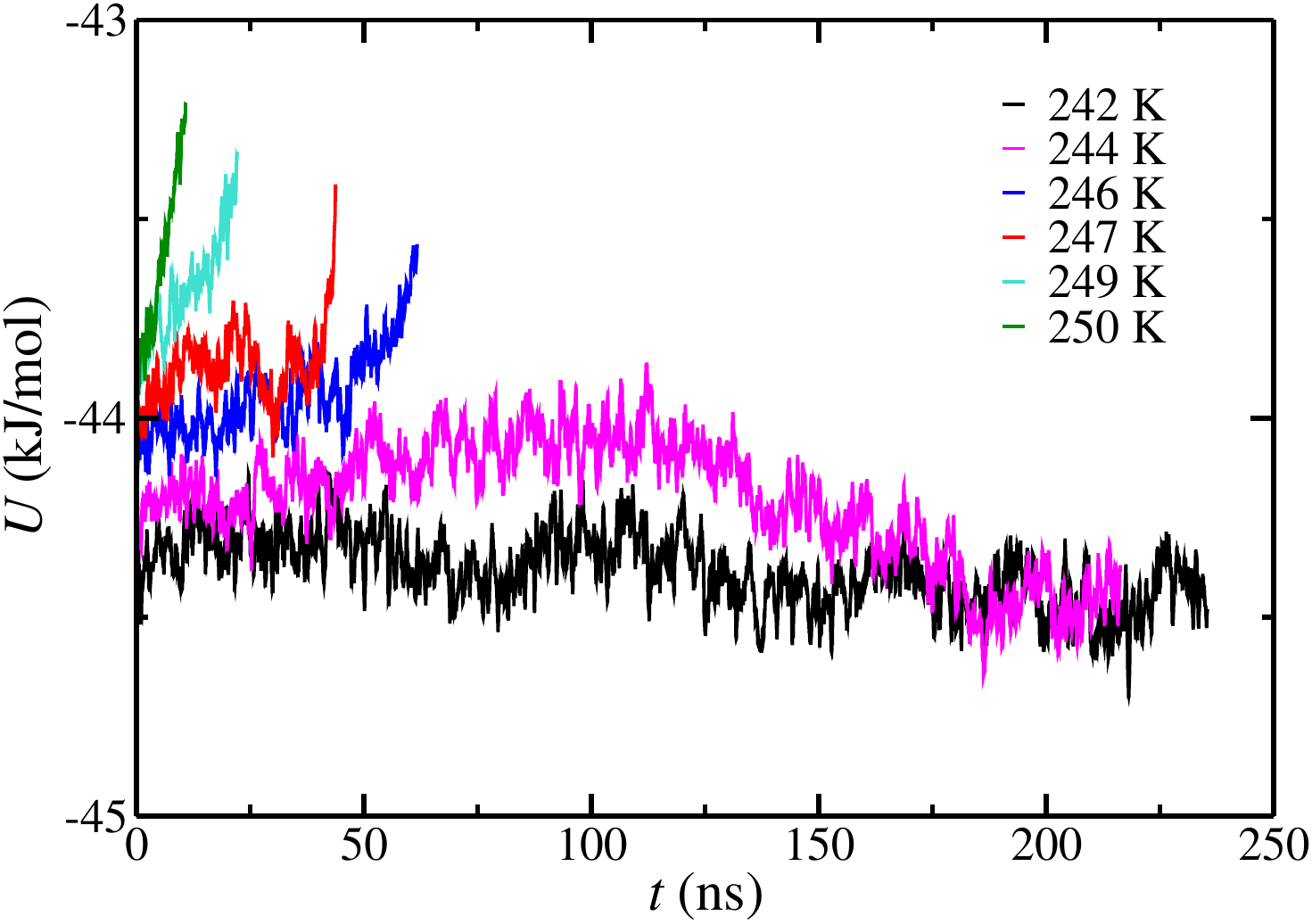}
\caption{Potential energy as a function of time as obtained from molecular dynamic simulations at different temperatures at $400\,\text{bar}$ using configuration S simulation box. In all cases, the initial aqueous phase is a pure water phase without NaCl.}
\label{CO2_simulations_conf2}
\end{figure}

Also, we determine the $T_3$ value at $400\,\text{bar}$ using configuration S as shown in Fig.~\ref{CO2_simulations_conf2}. The $T_3$ predicted using configuration S is $245(2)\,\text{K}$. As can be observed, there are differences between the $T_3$ values obtained using configurations L and S at $400\,\text{bar}$. It is important to recall that simulations have been carried out at the same conditions and using the same simulation details except for the simulation box size. It means that the difference in the $T_3$ value is caused by finite size effects and the inherited stochasticity of the direct-coexistence methodology. Also, it is interesting to compare this result with the results obtained in the work of Míguez \emph{et al.}~\cite{Miguez2015a} at $400\,\text{bar}$, $249(2)\,\text{K}$. The system size used in configuration S is closer to that used in the original work of Míguez \emph{et al.}~\cite{Miguez2015a} than the system size used in configuration L. However, the result obtained by using configuration S is different from that reported in the work of Míguez \emph{et al.}~\cite{Miguez2015a} As it has been pointed out previously, the cutoff employed in both works is different and this explains why the results obtained in both works are different even when the system sizes are similar. 

\begin{figure}
\centering
\includegraphics[width=0.9\columnwidth]{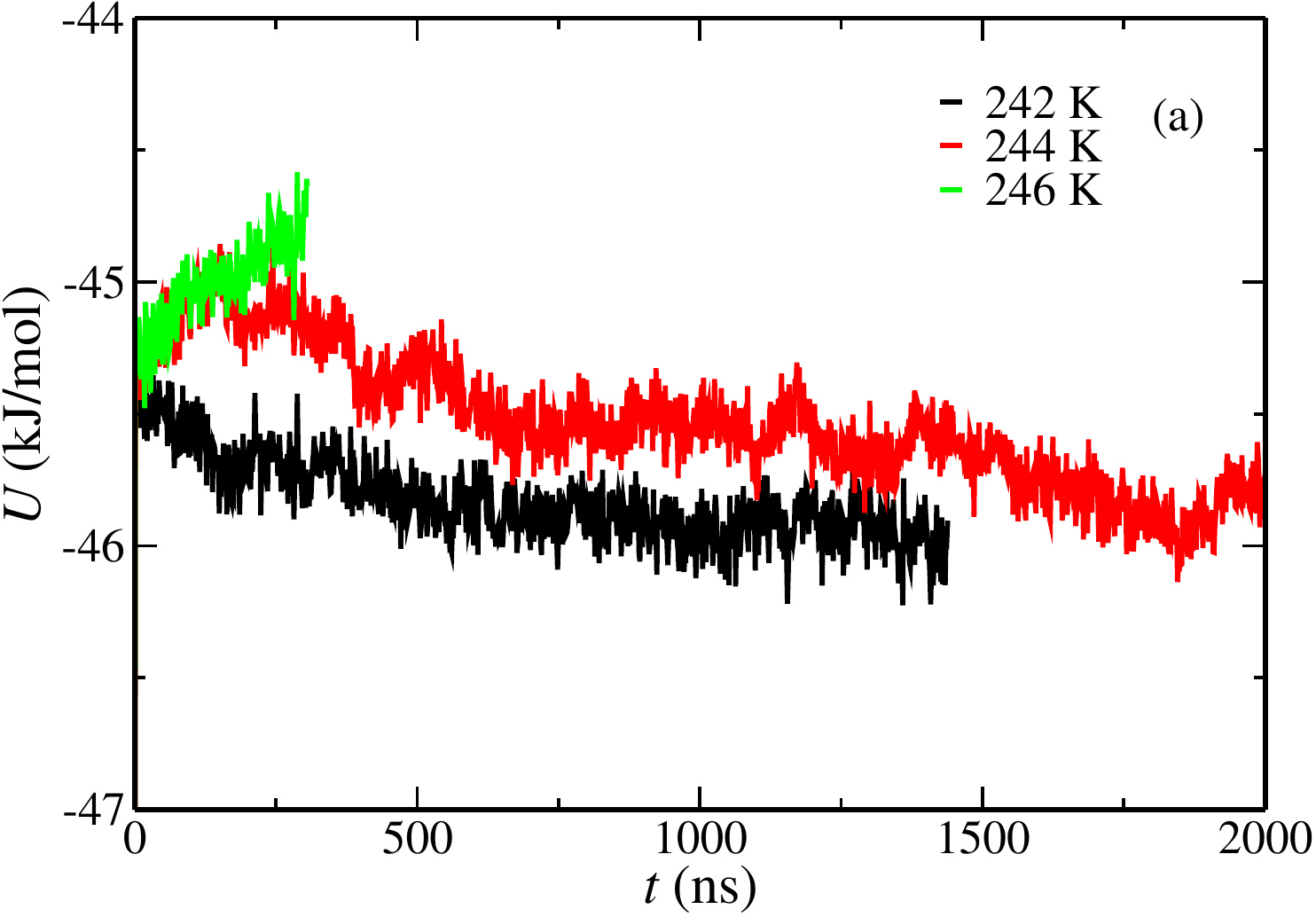}
\includegraphics[width=0.9\columnwidth]{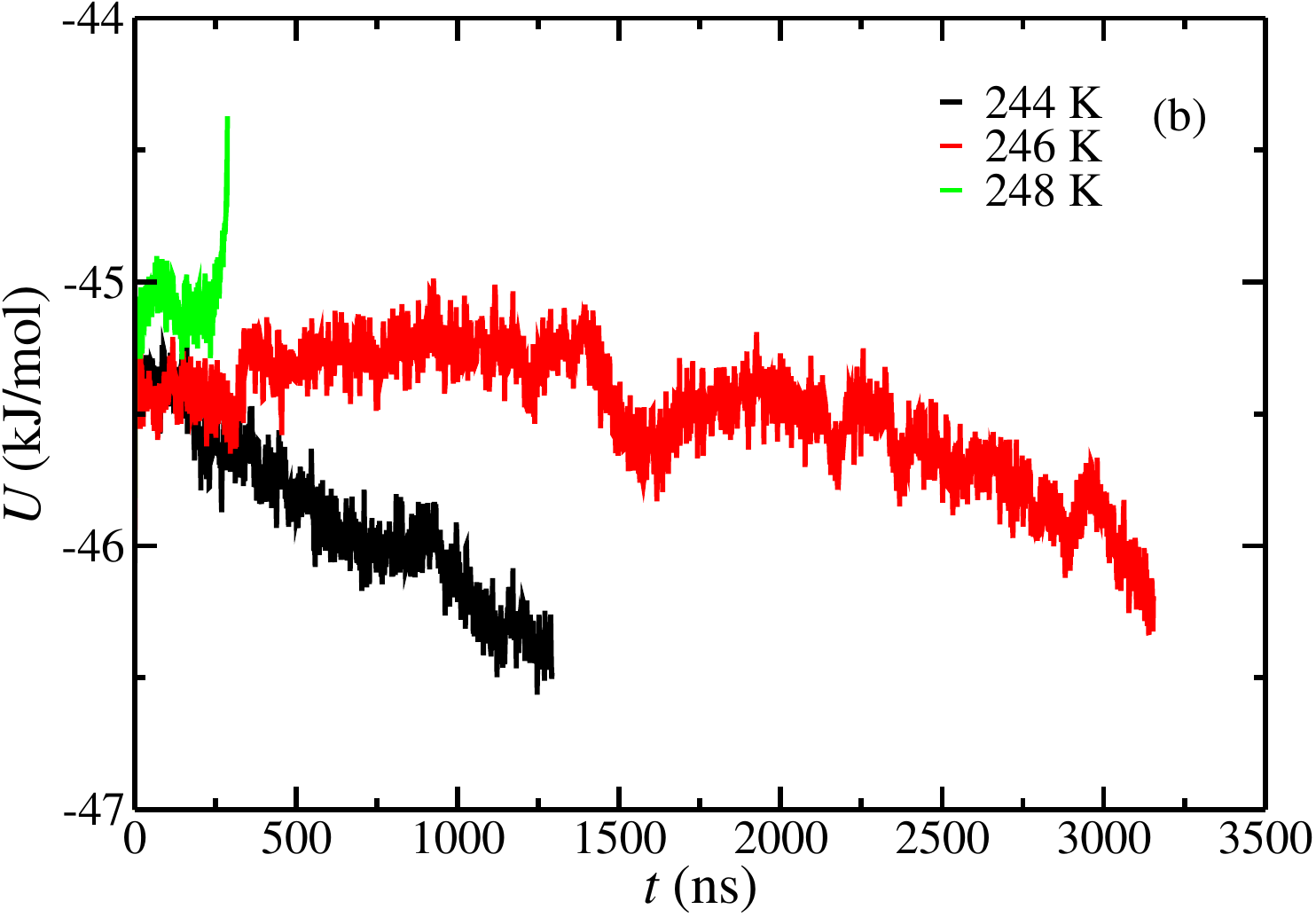}
\includegraphics[width=0.9\columnwidth]{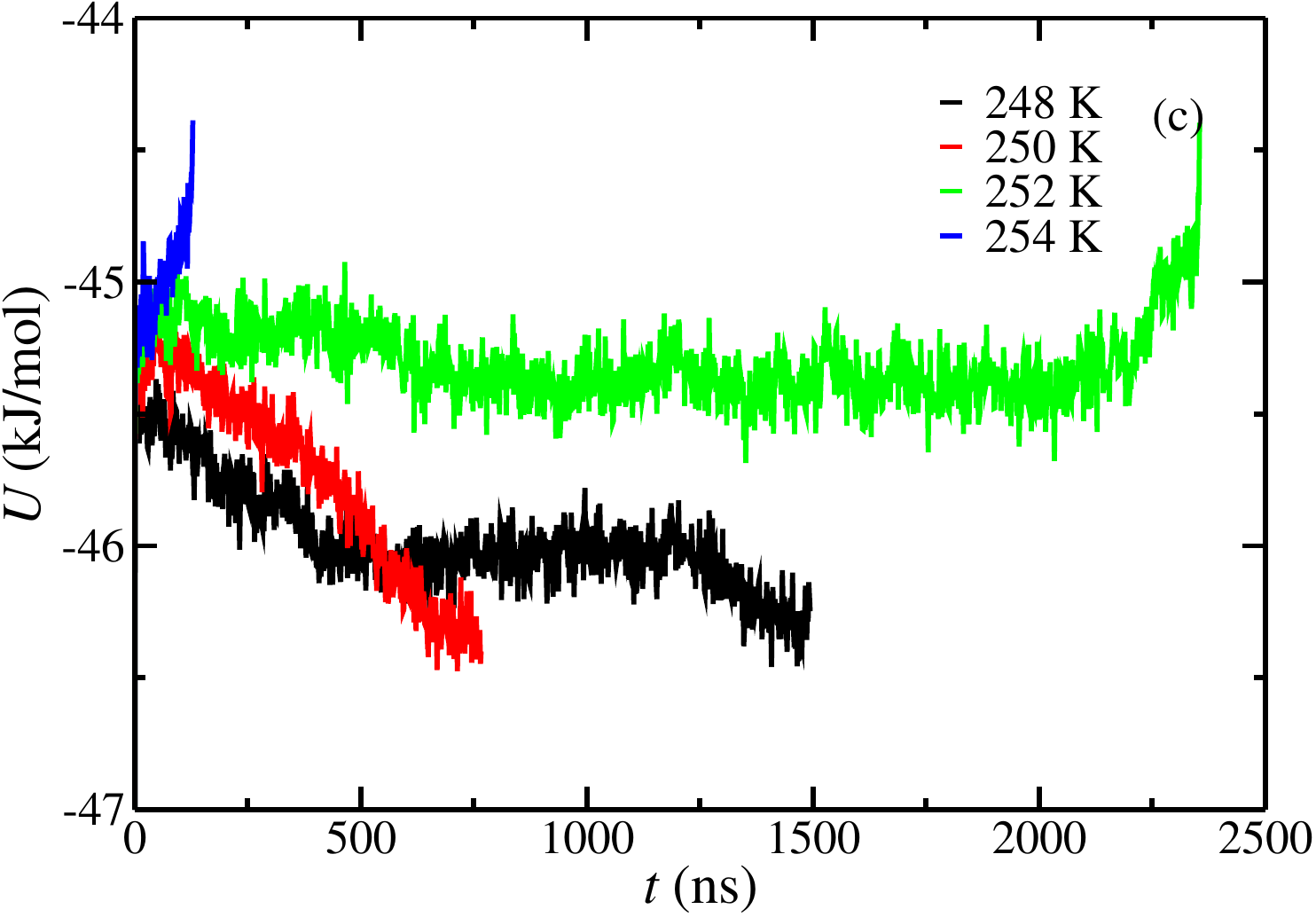}
\caption{Potential energy as a function of time as obtained from molecular dynamic simulations at different temperatures and pressures using configuration L simulation box. Figures (a), (b), and (c) correspond to results obtained at 100, 400, and $1000\,\text{bar}$ respectively. In all cases, the initial aqueous phase has a $0.6\,\text{m}$ NaCl concentration. The different temperatures are represented in the legend.}
\label{CO2_simulations_0.6}
\end{figure}

\subsection{NaCl 0.6\,m}
Now, we study the effect of a 0.6\,m NaCl concentration in the aqueous phase on the $T_3$ values obtained at 100, 400, and $1000\,\text{bar}$ using configuration L. The 0.6\,m NaCl concentration is especially interesting since it corresponds to the natural NaCl concentration of the seas and ocean. Most hydrate studies are carried out without taking into account the presence of NaCl in the aqueous phase. However, hydrates are common on the seabed, and for this reason, is essential to understand how the stability conditions of these compounds are modified by the presence of NaCl. The potential energy as a function of the simulation time obtained from configuration L at 100, 400, and $1000\,\text{bar}$ are represented in Fig. \ref{CO2_simulations_0.6}. The $T_3$ values obtained using configuration L at 100 (a), 400 (b), and 1000 (c) bar are 245(1), 247(1), and $251(1)\,\text{K}$ respectively. A summary of the $T_3$ results is presented in Table \ref{T3}. As can be observed in Fig. \ref{CO2_simulations_0.6} and Table \ref{T3},  the $T_3$ value is systematically reduced by $2\,\text{K}$ at each pressure considered in this work when the NaCl concentration is increased from 0.0 to $0.6\,\text{m}$. This is the expected behavior since the presence of NaCl in the aqueous phase decreases the $T_3$ of the hydrates.

It is also interesting to analyze the behavior of the dynamics of the system as a function of the NaCl concentration. As can be observed in Figs. \ref{CO2_simulations_sinsal} and \ref{CO2_simulations_0.6}, simulation times become larger when the NaCl concentration is increased. Yagasaki \emph{et al.}\cite{Yagasaki2014a} reported a similar slowing down effect at a NaCl $0.6\,\text{m}$ concentration for the CH$_4$ hydrate dissociation case. In general, we can conclude that simulation times when the NaCl concentration is $0.6\,\text{m}$ are about $1.5-2.0$ larger than those required in the absence of NaCl. It is also interesting to remark that larger hydrate systems required, in general, larger simulation times.~\cite{Blazquez2024b,Algaba2024a} This is important since an increment in the size of the system yields larger simulation times. Also, simulations become more expensive when the size of the system is increased. Finally, as has been pointed out previously,  the presence of NaCl in the aqueous phase affects the dynamic of the hydrates, and larger simulation times are required. This phenomenon is due to the fact that the salt acts as an inhibitor of the formation of the hydrate. The Na$^+$ cations and Cl$^-$ anions solvate the molecules of water, forming a hydration sphere around the ions, resulting in a reduction of the water molecules available to form the CO$_2$ hydrate structure under crystallization conditions ($T<T_3$). Another reason is that the solubility of CO$_2$ in the aqueous phase decreases when the salinity is increased.~\cite{Blazquez2024a,Sun2016a} In the work of Sun \emph{et al.},~\cite{Sun2016a} they calculated the solubility of CO$_2$ in water and in saline solutions in equilibrium with CO$_2$ hydrates. In this study, it is observed that the solubility of CO$_2$ in water decreases as the concentration of NaCl in the aqueous phase increases. This has a double effect, on the one hand, under dissociation conditions, the CO$_2$ releases from the dissociation of the hydrate can not pass through the aqueous solution to reach the CO$_2$ phase, leading to the formation of the hydrate again and delaying the dissociation of the hydrate. On the other hand,  this phenomenon also increases simulation times because the low solubility of CO$_2$ in saline solutions makes it difficult for the CO$_2$ molecules to pass from the CO$_2$ phase to the hydrate surface (through the aqueous phase), delaying the formation of new layers of CO$_2$ hydrate. In summary, one could be tempted to use smaller and more affordable system sizes to study how the presence of NaCl affects the dissociation temperature of hydrates. However, small systems are affected by finite-size effects and can not be used to determine accurately the dissociation temperature of these systems.~\cite{Blazquez2024b,Algaba2024a}

\subsection{NaCl $1.85\,\text{m}$}

\begin{figure}
\centering
\includegraphics[width=0.9\columnwidth]{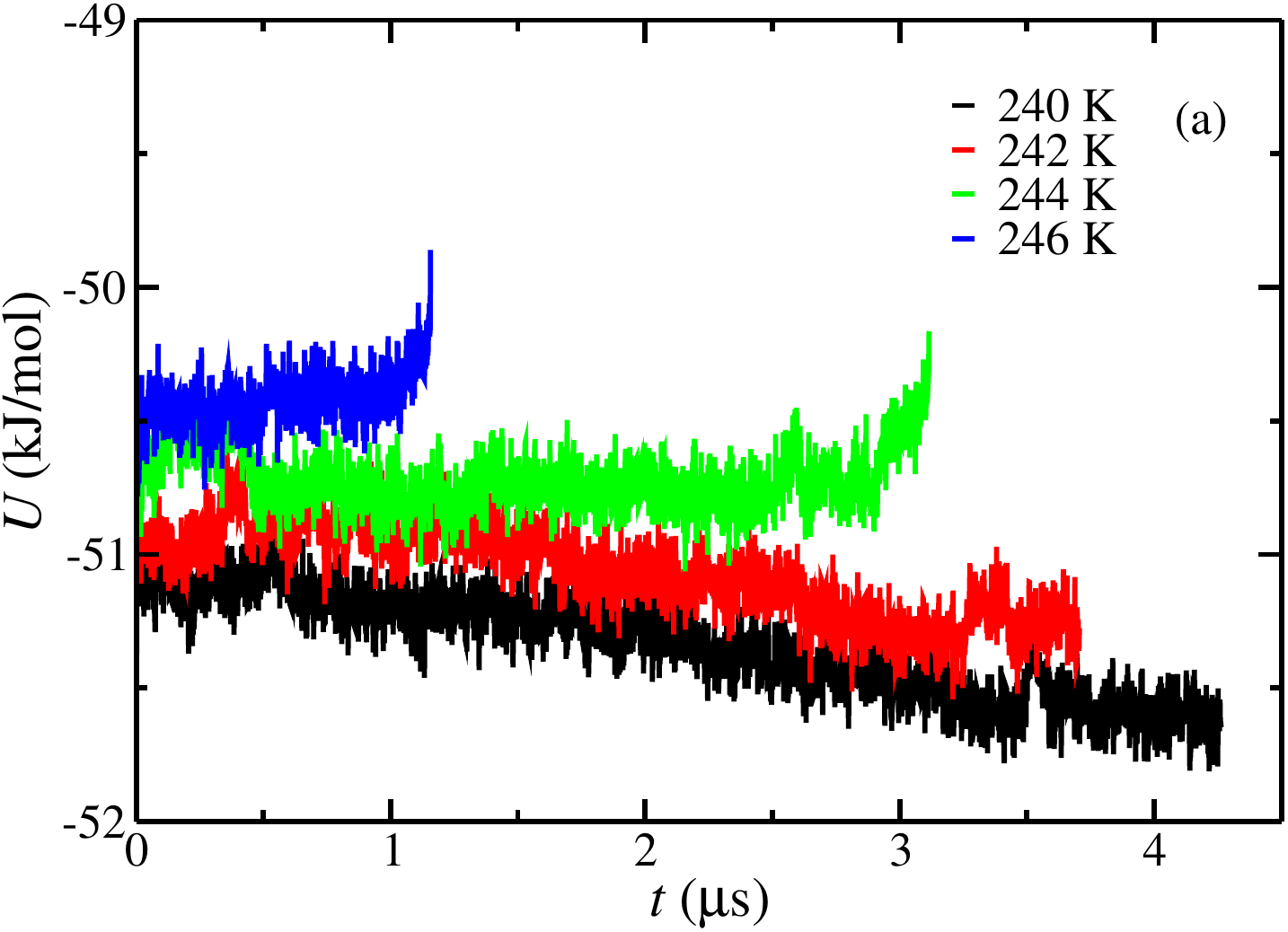}
\includegraphics[width=0.9\columnwidth]{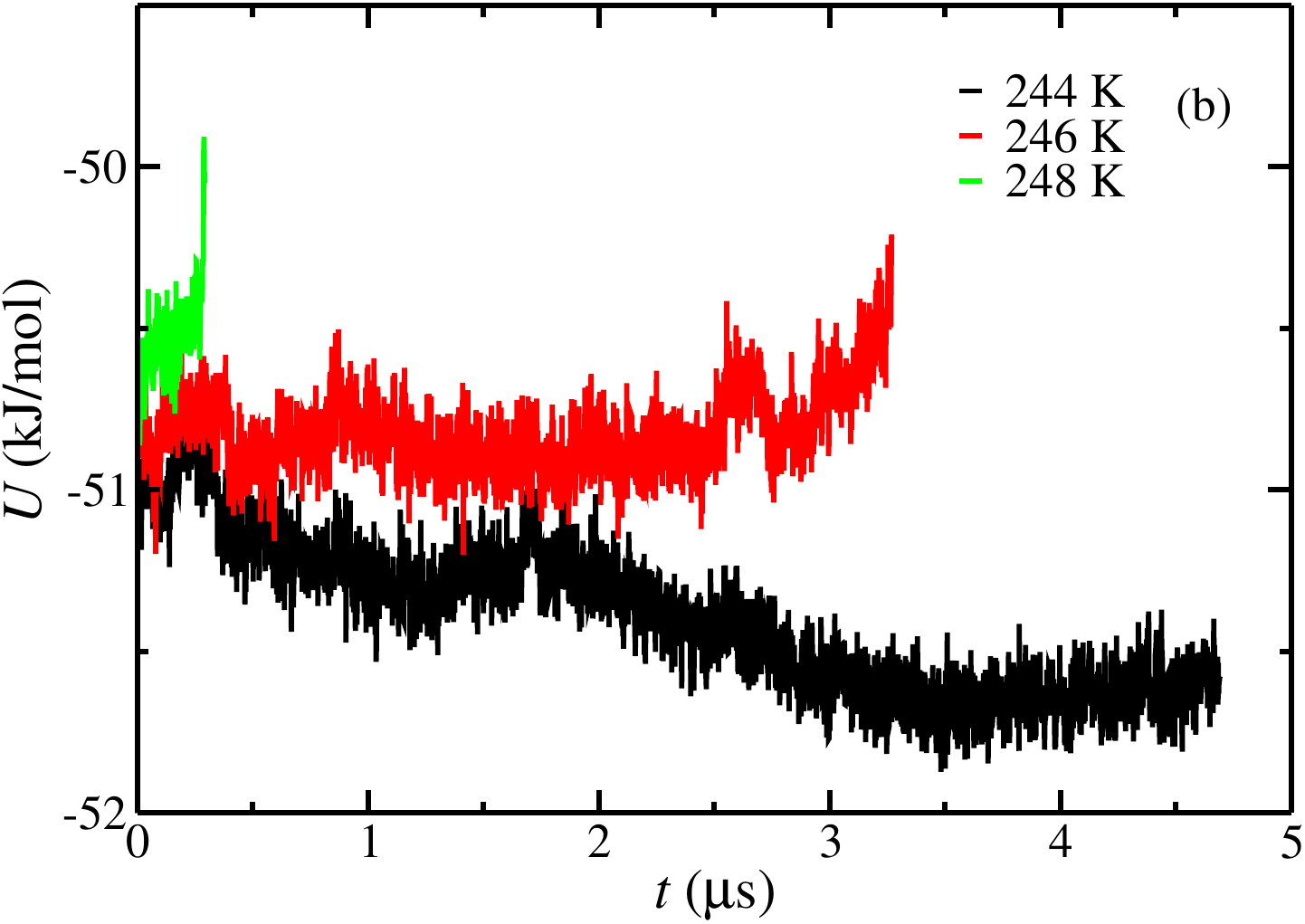}
\includegraphics[width=0.9\columnwidth]{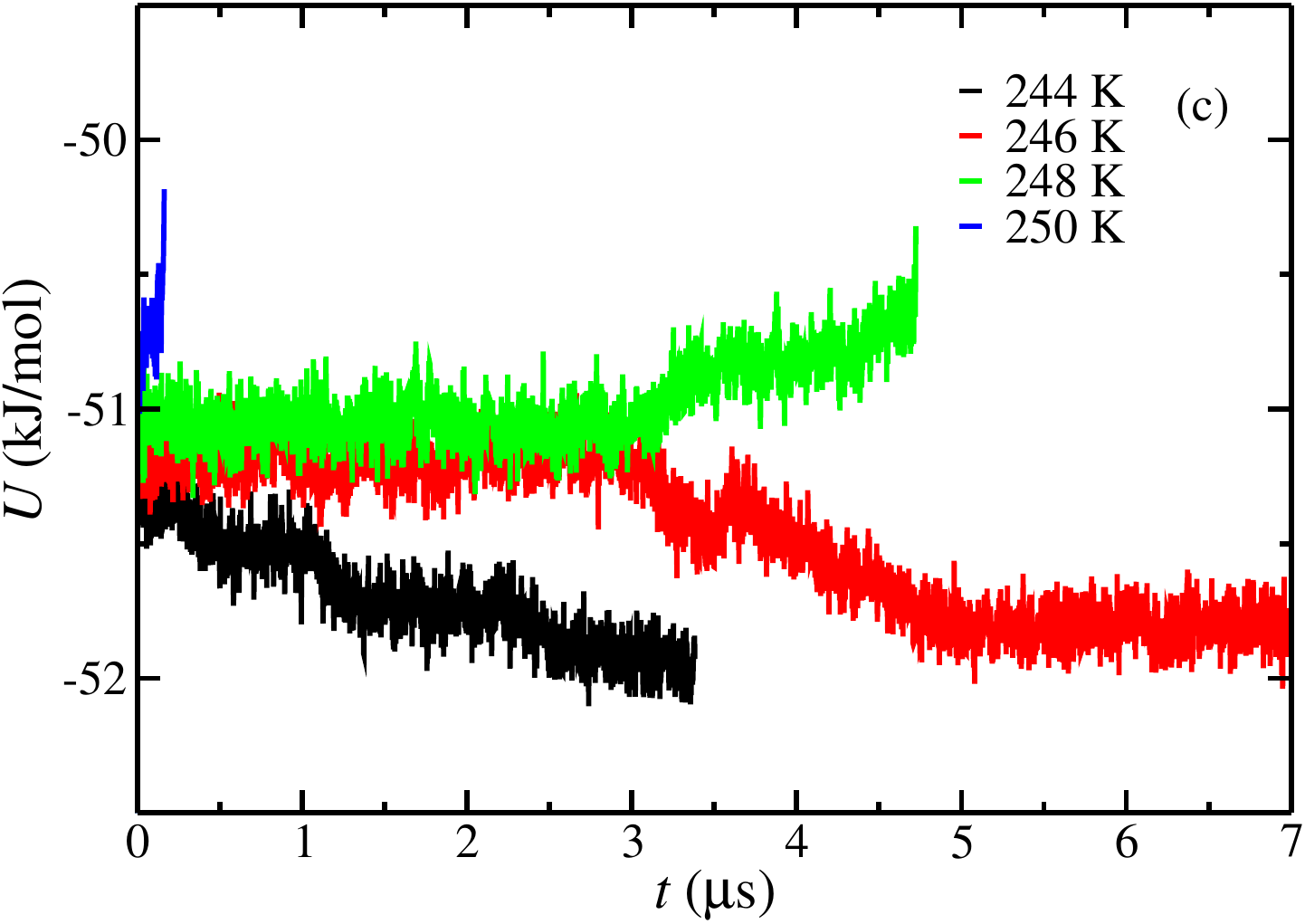}
\caption{Potential energy as a function of time as obtained from molecular dynamic simulations at different temperatures and pressures using configuration L simulation box. Figures (a), (b), and (c) correspond to results obtained at 100, 400, and $1000\,\text{bar}$ respectively. In all cases, the initial aqueous phase has a $1.85\,\text{m}$ NaCl concentration. The different temperatures are represented in the legend.}
\label{CO2_simulations_1.85}
\end{figure}

In this section, we determine the three-phase dissociation temperature, $T_3$, of the CO$_2$ hydrate when the initial aqueous phase has a $1.85\,\text{m}$ concentration of NaCl. We have analyzed this system using configuration L. As we have shown previously, the amount of NaCl in the aqueous solution has an effect on the $T_3$ as well as on the simulation time required by the system to evolve. As can be observed in Fig. \ref{CO2_simulations_1.85} and in Table \ref{T3}, the dissociation temperature as a function of the NaCl concentration behaves as expected, when the NaCl concentration is increased, the $T_3$ value is decreased. In particular, $T_3$ values at a NaCl concentration of $1.85\,\text{m}$ are 243(1), 245(1), and $247(1)\,\text{K}$ at 100, 400 and $1000\,\text{bar}$ respectively. At 100 and $400\,\text{bar}$, the $T_3$ decreases $2\,\text{K}$ when the NaCl concentration is increased from 0.0 to $0.6\,\text{m}$, and when it is increased from 0.6 to $1.85\,\text{m}$. At $1000\,\text{bar}$, the $T_3$ also decreases when the NaCl concentration is increased. As at 100 and $400\,\text{bar}$, the $T_3$ value at $1000\,\text{bar}$ decreases $2\,\text{K}$ when the NaCl concentration is increased from 0.0 to $0.6\,\text{m}$. However, at $1000\,\text{bar}$, when the NaCl concentration is increased from 0.6 to $1.85\,\text{m}$, the $T_3$ value decreases $4\,\text{K}$, from $251(1)\,\text{K}$ at $0.6\,\text{m}$ to $247(1)\,\text{K}$ at $1.85\,\text{m}$.

It is also interesting to analyze the dynamic of the system at $1.85\,\text{m}$ of NaCl. As has been pointed out previously, when the amount of NaCl in the aqueous phase is increased, the time required by the hydrate to grow or melt is also increased. In particular, simulation times, from 0.6 to $1.85\,\text{m}$, have been increased by a $1.5-2.5$ factor. It means that simulation times when the initial aqueous phase has $1.85\,\text{m}$ NaCl concentration are about $3-5$ larger than when there is no NaCl in the aqueous phase. Notice that when the temperatures are close to the $T_3$ values, the simulation time required by the system to evolve is even larger. The amount of NaCl in the aqueous phase not only affects the time required by the hydrate to grow but also to melt. The melting process used to be faster than the growth of the hydrate phase. However, the presence of NaCl slows both processes. As a consequence, it is necessary to increase drastically the simulation time in order to study correctly the behaviour of these systems.

 \subsection{NaCl $3.0\,\text{m}$}

\begin{figure}
\centering
\includegraphics[width=0.9\columnwidth]{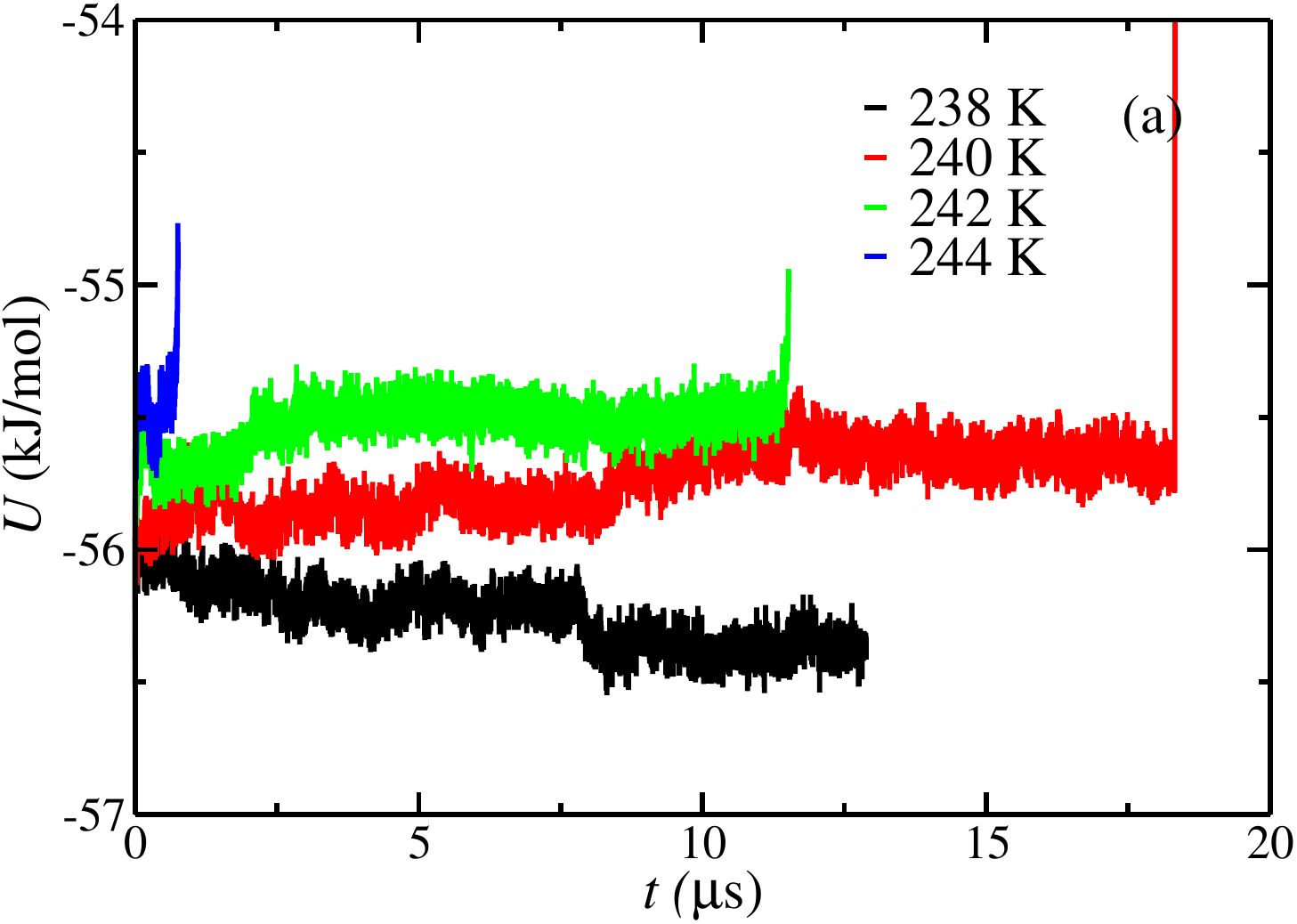}
\includegraphics[width=0.9\columnwidth]{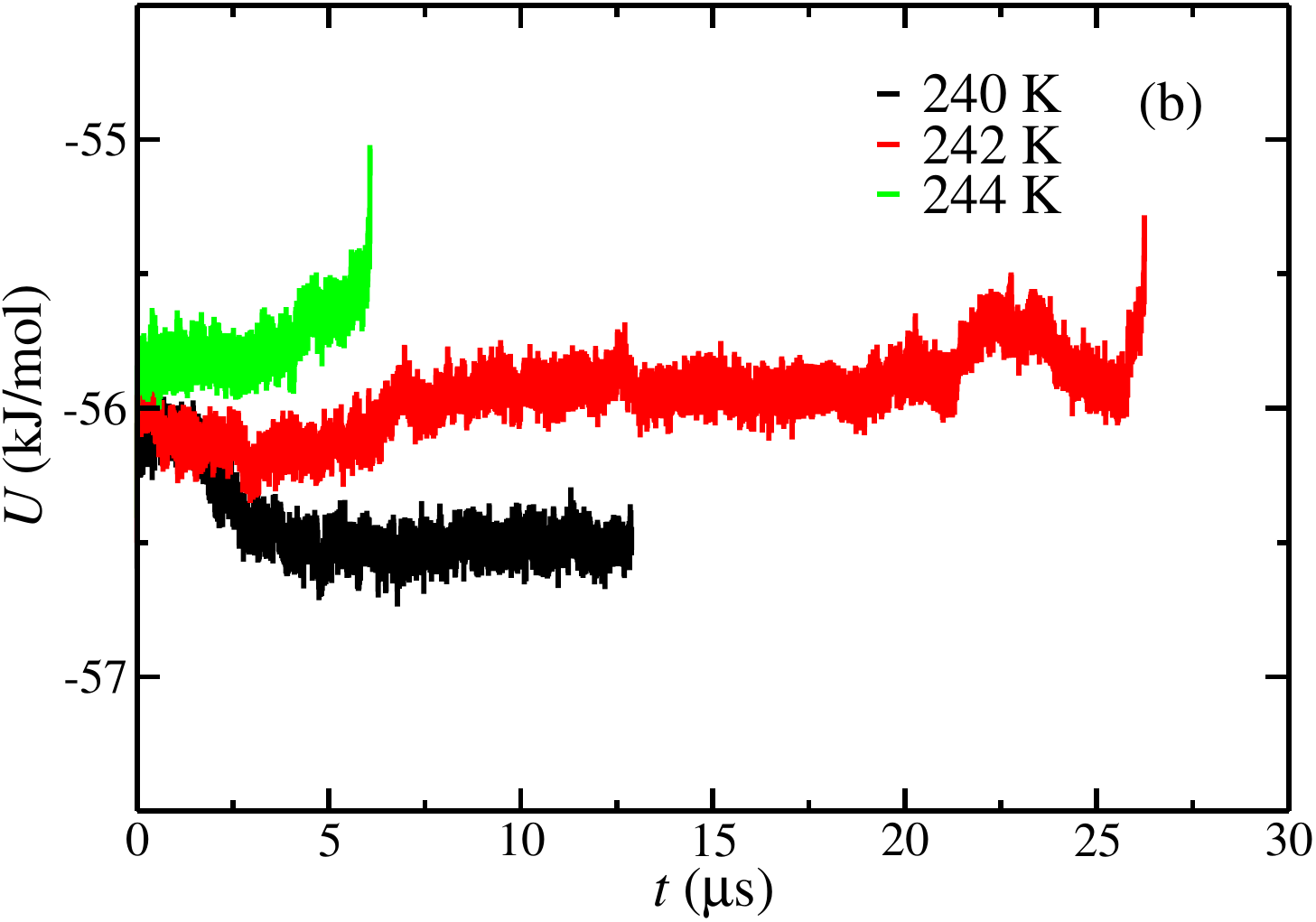}
\includegraphics[width=0.9\columnwidth]{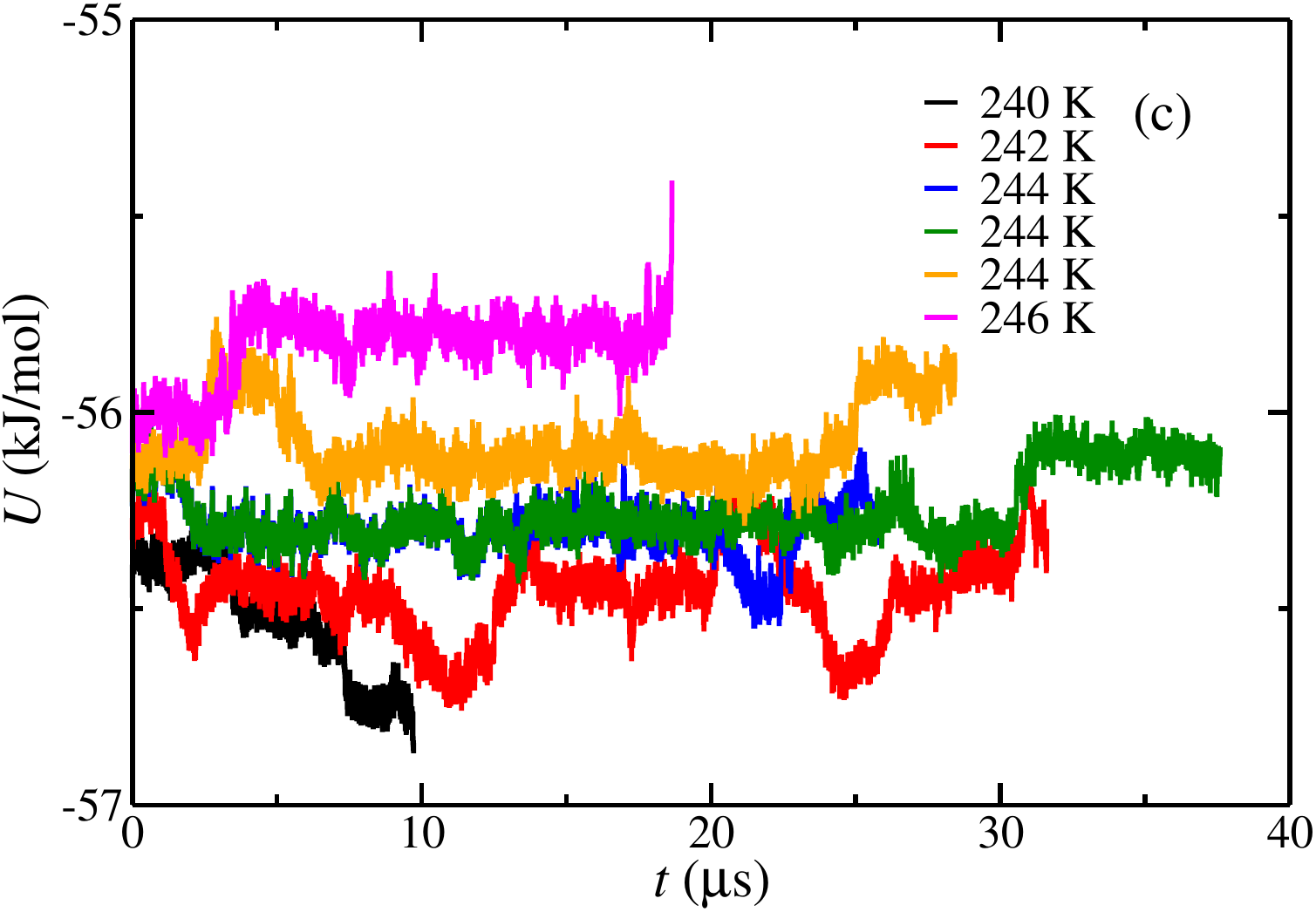}
\caption{Potential energy as a function of time as obtained from molecular dynamic simulations at different temperatures and pressures using configuration L simulation box. Figures (a), (b), and (c) correspond to results obtained at 100, 400, and $1000\,\text{bar}$ respectively. In all cases, the initial aqueous phase has a 3.0 m NaCl concentration. The different temperatures are represented in the legend.}
\label{CO2_simulations_3.0}
\end{figure}

Finally, we determine the three-phase dissociation temperature, $T_3$, of the CO$_2$ hydrate when the initial aqueous phase has a $3.0\,\text{m}$ concentration of NaCl. We study the cryoscopic decrease effect at this concentration using configuration L at 100, 400, and $1000\,\text{bar}$. Also, we study the effect of the NaCl concentration at $400\,\text{bar}$ using configuration S. First, we focus on the results obtained using configuration L. Figure \ref{CO2_simulations_3.0} shows the potential energy as a function of the simulation time at 100 (a), 400 (b), and 1000 (c) bar and different temperatures. As expected, the increase of the salinity of the aqueous phase reduces the $T_3$ values obtained at each pressure. In particular, the $T_3$ values obtained at 100, 400, and $1000\,\text{bar}$ when the initial aqueous phase has a $3.0\,\text{m}$ concentration of NaCl are 239(2), 242(2), and $244(2)\,\text{K}$ respectively. If we compare these results with those obtained without the presence of NaCl ($0.0\,\text{m}$), we find that $T_3$ values are reduced by 8, 7, and $9\,\text{K}$ at 100, 400, and $1000\,\text{bar}$, respectively. This is in good agreement with the cryoscopic decrease effect reported in the literature.~\cite{Duan2006,Dholabhai1993a} A further comparison with experimental data is presented in the next Section.

\begin{figure}
\centering
\includegraphics[width=1.0\columnwidth]{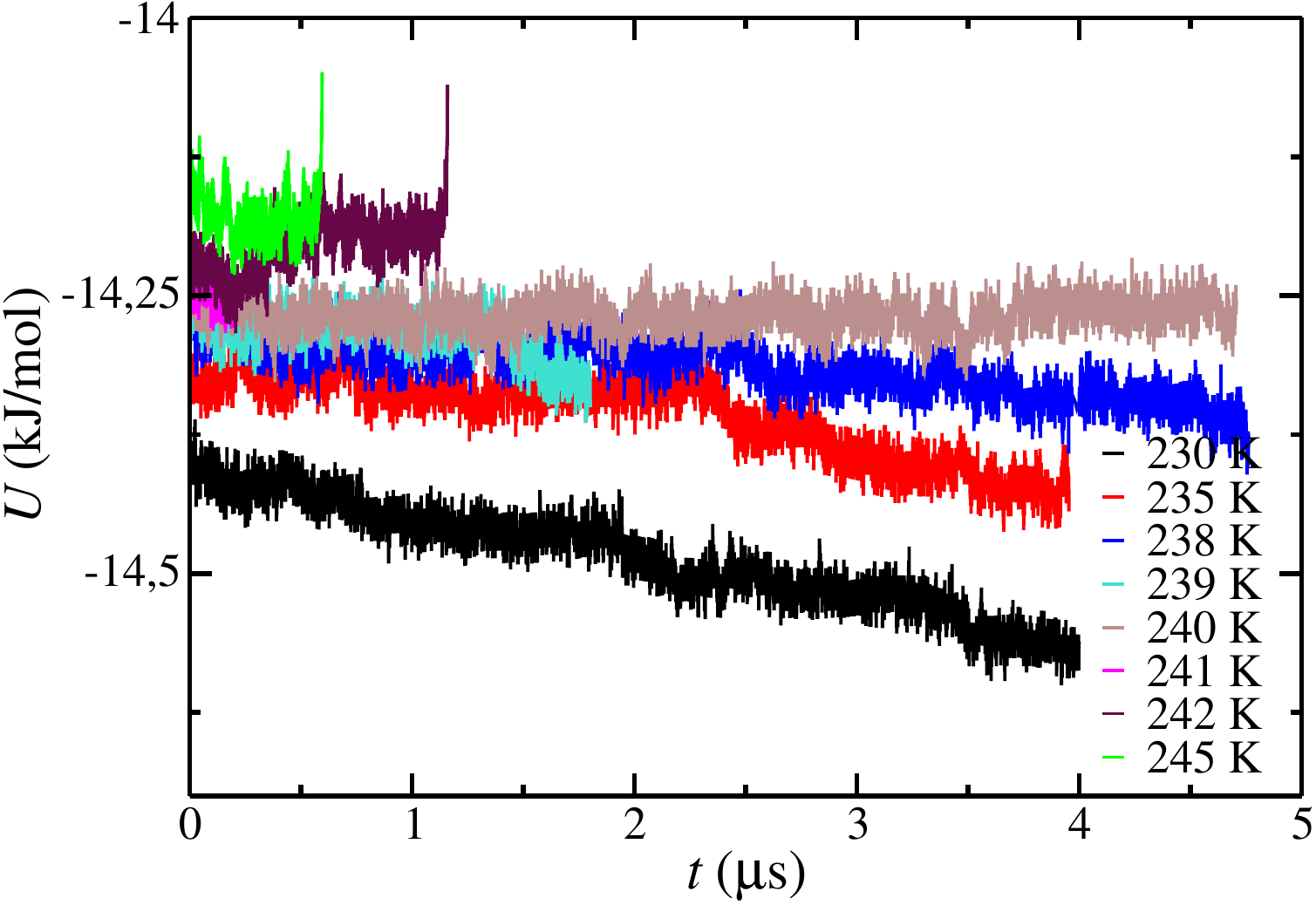}
\caption{Potential energy as a function of time as obtained from molecular dynamic simulations at different temperatures at $400\,\text{bar}$ using configuration S simulation box. In all cases, the initial NaCl concentration in the aqueous phase is $3.0\,\text{m}$.}
\label{CO2_conf2_3m}
\end{figure}

We also studied the cryoscopic decrease effect on the CO$_2$ hydrate at 400\,bar using configuration S. As can be seen in Fig. \ref{CO2_conf2_3m}, the $T_3$ value obtained using configuration S is $240(1)\,\text{K}$. Although the values obtained, at NaCl 3.0\,m, using configuration L and S are the same within the error bars, the results obtained without the presence of NaCl, at NaCl $0.0\,\text{m}$, are different, 249(1) and $245(1)\,\text{K}$  respectively. As a consequence, the predicted cryoscopic decrease effect, i.e., the reduction of the $T_3$ value as a function of the NaCl concentration, is different using configurations L and S. In particular, the $T_3$ decreases by 8 and $5\,\text{K}$  using configurations L and S respectively. It means that at $400\,\text{bar}$, the cryoscopic decrease effect
obtained from configurations L and S shows a difference of $3\,\text{K}$. Although a $3\,\text{K}$  difference could seem negligible, it is important to notice that $3\,\text{K}$ represents the 60$\%$ of the total cryoscopic decrease value obtained from configuration S.

Finally, we analyze the dynamic of the system with an initial 3.0\,m NaCl concentration in the aqueous phase. As it is shown in Fig. \ref{CO2_simulations_3.0}, the simulation times required to observe the growing/melting behavior of the system become extremely larger. In fact, we do not observe any concluding behavior at $1000\,\text{bar}$ and $T=244\,\text{K}$ after running three different seeds for at least $25\mu s$. It means that simulation times required at $3.0\,\text{m}$ becomes 3-5 times larger than at $1.85\,\text{m}$ and, hence, 10-20 times larger than that at $0.0\,\text{m}$. Here it is important to remark that the dynamic of the system becomes slower not only because of the increment of NaCl but also because simulation temperatures become lower due to the cryoscopic decrease effect.

\begin{table}
\caption{$T_3$ (K) values as a function of the pressure and the NaCl concentration (m) as obtained from molecular dynamic simulations using configuration L.}
\label{T3}
\resizebox{8cm}{!} {
\begin{tabular}{c c c c c c c c c c c c c c c c c c c c c}
\hline
\hline
\multirow{2}{*}{Pressure (bar)}& & & \multicolumn{11}{c}{Temperature (K)}\\
          \cline{3-4} \cline{6-7} \cline{9-10} \cline{12-13}
& & \multicolumn{2}{c}{0.0\,m} & &
\multicolumn{2}{c}{0.6\,m} & &
\multicolumn{2}{c}{1.85\,m} & &
\multicolumn{2}{c}{3.0\,m} & \\
     
      \hline
      100 & & \multicolumn{2}{c}{247(1)} & &\multicolumn{2}{c}{245(1)} & & \multicolumn{2}{c}{243(1)} & & \multicolumn{2}{c}{239(1)} \\
      400 & & \multicolumn{2}{c}{249(1)} & & \multicolumn{2}{c}{247(1)} & & \multicolumn{2}{c}{245(1)} & & \multicolumn{2}{c}{241(1)} \\
      1000 & & \multicolumn{2}{c}{253(1)} & &\multicolumn{2}{c}{251(1)} & & \multicolumn{2}{c}{247(1)} & & \multicolumn{2}{c}{244(2)} \\
      \hline
      \hline
\end{tabular}
}
\end{table}

\section{Cryoscopic decrease effect}

Finally, we analyze the cryoscopic decrease effect predicted by molecular dynamic simulation. As has been explained previously, the water, TIP4P/2005,~\cite{Abascal2005a} and CO$_2$, TraPPE,~\cite{Potoff2001a}  molecular models employed in this work are able to reproduce qualitatively the dissociation temperature of the CO$_2$ hydrate. The results obtained using these two models underestimate the dissociation temperature by $\approx30-45\,\text{K}$. However, in this work, we are interested in the cryoscopic decrease effect, i.e., how the $T_3$ value is reduced as a function of the salinity. The total temperature reduction, $\Delta T$, as a function of the NaCl concentration, can be expressed as the difference between the $T_3$ value when the initial aqueous solution is pure water minus the $T_3$ value when the initial aqueous phase contains NaCl at a certain concentration. Notice that when the cryoscopic decrease effect is expressed as $\Delta T$, the initial $T_3$ value becomes irrelevant and allows us to analyze whether the methodology, as well as the system sizes and models used in this work, are able to reproduce the experimental data.  

\begin{figure}
\centering
\includegraphics[width=1.0\columnwidth]{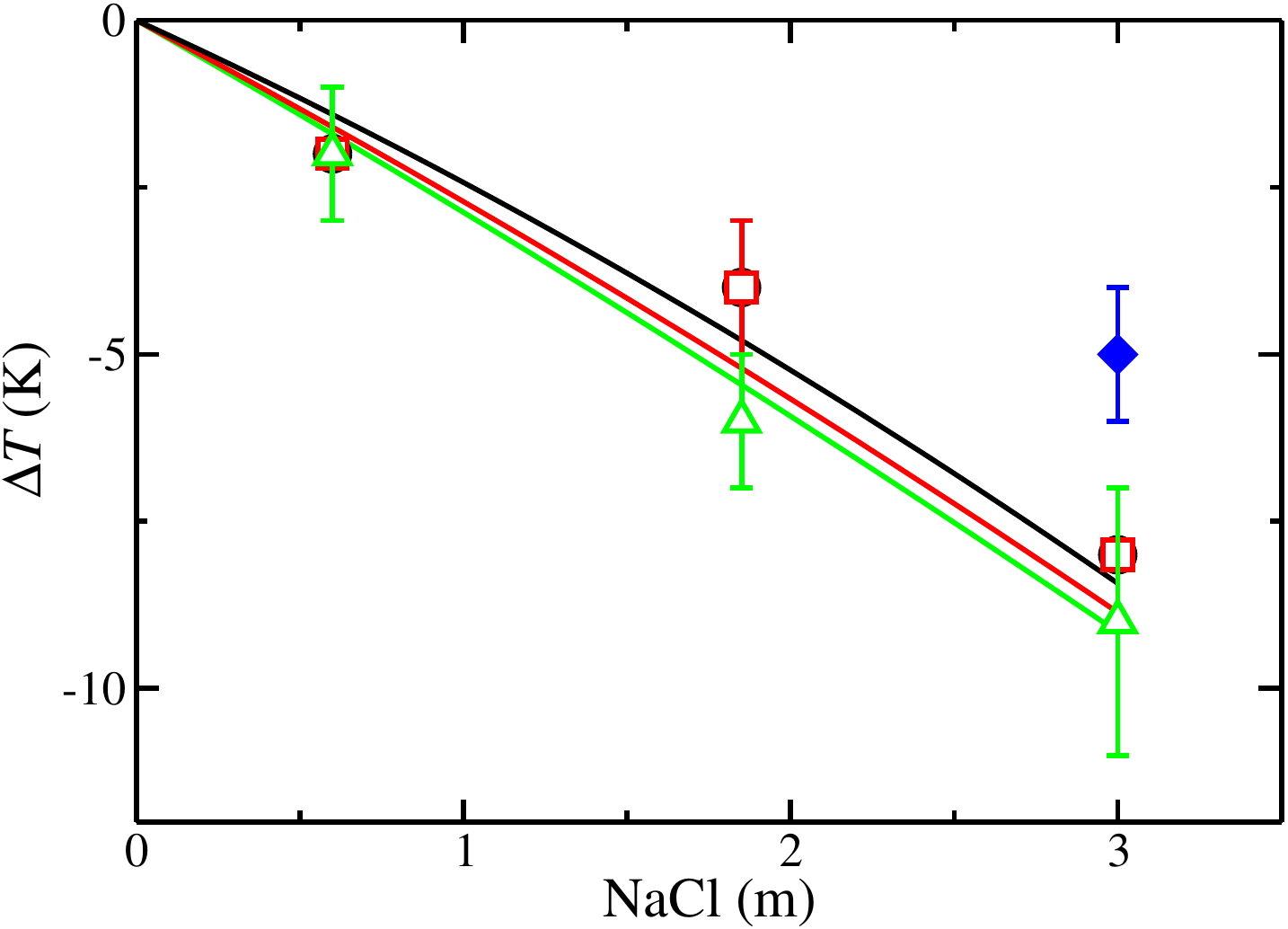}
\caption{Cryoscopic decrease effect as a function of the NaCl molality. Black, red, and green lines correspond to the experimental cryoscopic decrease effect for the CO$_2$ hydrate at 100, 400, and $1000\,\text{bar}$, respectively.~\cite{Duan2006} Open black circles, red squares, and green up-triangles correspond to the results obtained in this work using configuration L at 100, 400, and $1000\,\text{bar}$ respectively. The filled blue diamond corresponds to the result obtained at $400\,\text{bar}$ using configuration S. Notice that black and red symbols (100 and $400\,\text{bar}$) overlap in all cases.}
\label{Cryo-effect}
\end{figure}

Figure \ref{Cryo-effect} shows a representation of the cryoscopic decrease effect a as function of the NaCl molality in the aqueous phase. As can be observed, the cryoscopic decrease effect is not a linear function of the molality. The $T_3$ value shows a poor dependency with the NaCl concentration at low salinity values. However, the slope becomes sharper when the NaCl molality is increased. It is also interesting to analyze the effect of the pressure on the cryoscopic decrease effect. The three pressures studied in this work present the same qualitative behavior. Also, it is interesting to observe that the cryoscopic decrease effect slightly increases when the pressure is increased.

As has been explained previously, the TIP4P/2005\cite{Abascal2005a} and the TraPPE\cite{Potoff2001a} models for water and CO$_2$ have been previously employed to describe the dissociation line of the CO$_2$ hydrate.~\cite{Miguez2015a} However, this is the first time that the Madrid-2019 NaCl model\cite{Zeron2019} has been employed to describe the dissociation line of a hydrate at different salinity concentrations. As can be observed, the results obtained in this work using configuration L are in excellent agreement with the experimental results taken from the literature. Also, we can observe that the result obtained by using configuration S underestimates the cryoscopic decrease effect at a NaCl concentration of $3.0\,\text{m}$. Here, it is important to remark that the size of the configuration L system is far from being arbitrary. As it has been demonstrated by some of us in a previous series of papers,~\cite{Blazquez2024b,Algaba2024a,Algaba2024b} the size of the simulation box has to be chosen carefully in order to avoid finite-size effects on the $T_3$ determination. According to our previous works,~\cite{Blazquez2024b,Algaba2024a} configuration L is large enough and contains enough molecules to avoid such finite-size effects. However, configuration S should be use carefully since it is too small and the results provided by this configuration are affected by finite-size effects.

\section{Conclusions}

In this work, we have used the direct coexistence simulation technique to study the cryoscopic descent effect on the three-phase dissociation temperature of the CO$_{2}$ hydrate at different initial NaCl molalities in the aqueous phase and several pressures (100, 400, and $1000\,\text{bar}$). Notice that the number of data reported in this work is of the same order of previous studies reported in the literature where the cryoscopic descent effect of ice-like systems is studied through molecular simulation.\cite{Blazquez2023b,Fernandez-Fernandez2019a,Lamas2022a} The water and CO$_2$ molecules are modeled using the TIP4P/2005 and the TraPPE force fields, respectively. To describe the water-salt interactions in our system, we have used the Madrid-2019 force field\cite{Zeron2019,10.1063/5.0077716}. This force field has been developed for several salts in combination with the TIP4P/2005 water model. According to our previous work,~\cite{Blazquez2024b,Algaba2024a} the system L is large enough to avoid finite size effects. However, we have also performed the same study with a smaller configuration (S) at $400\,\text{bar}$ and at NaCl 0 and $3.0\,\text{m}$ to analyze finite size effects on the three-phase dissociation temperature as well as on the cryoscopic descent effect. 

In the absence of NaCl, the $T_3$ values obtained by configuration L are in good agreement with the data reported previously in the literature.~\cite{Miguez2015a} Contrary, configuration S underestimates the $T_3$ value at $400\,\text{bar}$ by $4\,\text{K}$. Taking into account that the determination of the $T_3$ is the first step in the cryoscopic descent effect study, a reasonable question arises: How much does the size of the system affect the cryoscopic descent? As we demonstrate in this work, the cryoscopic descent effect results at NaCl 0.6, 1.85, and $3.0\,\text{m}$ obtained with configuration L are in excellent agreement with the experimental data~\cite{Duan2006} reported in the literature in the range of pressure considered in this work. However, the result obtained by using configuration S at $400\,\text{bar}$ and NaCl $3.0\,\text{m}$ clearly underestimates the cryoscopic descent effect by $3\,\text{K}$ which represents the 60$\%$ of the total cryoscopic decrease value obtained from configuration S.

We also analyze the effect of pressure on the cryoscopic decrease effect using configuration L. As has been discussed in Fig. \ref{Cryo-effect}, the cryoscopic decrease obtained in this work from molecular dynamic simulations at 100 and $400\,\text{bar}$ is the same at the different NaCl concentrations considered in this study. At $1000\,\text{bar}$, the cryoscopic decrease effect is slightly higher than those obtained at 100 and $400\,\text{bar}$. The simulation results obtained in this work and the previously reported experimental data~\cite{Duan2006} are in excellent agreement within the uncertainties. This demonstrated that the combination of the Madrid-2019 force fields\cite{Zeron2019,10.1063/5.0077716} (including the TIP4P/2005 water model) and the CO$_2$ TraPPE model\cite{Potoff2001a} provide an accurate description of the cryoscopic decrease effect on the CO$_2$ hydrate as a function of the NaCl concentration in the range of pressures considered in this work.

Finally, we also analyze the dynamic of the system e.g., we analyze the simulation times required to observe if the hydrate phase grows or melts. We conclude that the dynamic of the system becomes extremely lower when the initial amount of NaCl in the aqueous phase is increased. This is provoked due to two effects. The first one is the concentration of NaCl in the aqueous phase, which probably decreases the water disponibility due to a solvation effect. The second one is because simulation
temperatures become lower due to the cryoscopic decrease effect. As a result of both effects, simulation times required in the absence of NaCl are $\approx10-20$ shorter than those required at NaCl $3.0\,\text{m}$. As a consequence, the simulations of this work added together, result in a total simulation time of about 300 $\mu$s. 

In summary, this work provides an accurate description of the $T_3$ and the cryoscopic decrease effect determination of the CO$_2$ hydrate at different pressures. We validate the use of the Madrid-2019 force fields (salt + TIP4P/2005 water models) for the study of the cryoscopic decrease effect in the particular case of hydrates. This finding contributes to the understanding of the stability conditions of hydrates under realistic seawater conditions and, together with the previous results obtained by some of the authors,~\cite{Blazquez2023b} leads the way to explore other hydrates using the Madrid-2019 force field.

\section*{Acknowledgements}

The authors extend their heartfelt appreciation to Dr. E. Dendy Sloan for his invaluable contributions to the study of gas hydrates. His groundbreaking work has been instrumental in advancing experimental research while also paving the way for molecular simulations, offering deeper insights into hydrate structure and behavior. His legacy remains a guiding force in both theoretical and applied research, inspiring future generations of scientists. We are profoundly grateful for his unwavering dedication and the enduring influence of his work in academia and industry. This work was funded by Ministerio de Ciencia e Innovaci\'on (Grants No.~PID2021-125081NB-I00 and PID2022-136919NB-C32), Junta de Andalucía (P20-00363), and Universidad de Huelva (P.O. FEDER UHU-1255522, FEDER-UHU-202034, and EPIT1282023), the first four were co-financed by EU FEDER funds. We greatly acknowledge RES resources at Picasso provided by The Supercomputing and Bioinnovation Center of the University of Malaga to FI-2024-1-0017. S.B. acknowledges Ayuntamiento de Madrid for a Residencia de Estudiantes grant. The authors gratefully acknowledge the Universidad Politecnica de Madrid (www.upm.es) for providing computing resources on Magerit Supercomputer. 

\section*{AUTHORS DECLARATIONS}

\section*{Conflicts of interest}

The authors have no conflicts to disclose.


\section*{Data availability}

The data that support the findings of this study are available within the article.


\bibliography{bibfjblas}

\end{document}